\documentclass[11pt,a4paper]{article}
\pdfoutput=1
\usepackage{jinstpub}
\usepackage{multirow}
\usepackage{enumitem}
\usepackage{lineno}
\usepackage{blindtext}
\usepackage[nottoc]{tocbibind}
\numberwithin{table}{section}
\usepackage{subcaption}
\usepackage[T1]{fontenc} 
\usepackage[latin1]{inputenc} 
\usepackage{graphicx}
\usepackage{natbib}
\usepackage{hyperref}
\usepackage{doi}
\usepackage{float}

\title{Design, upgrade and characterization of the silicon photomultiplier front-end for the AMIGA detector at the Pierre Auger Observatory}

\author{\includegraphics[height=30mm]{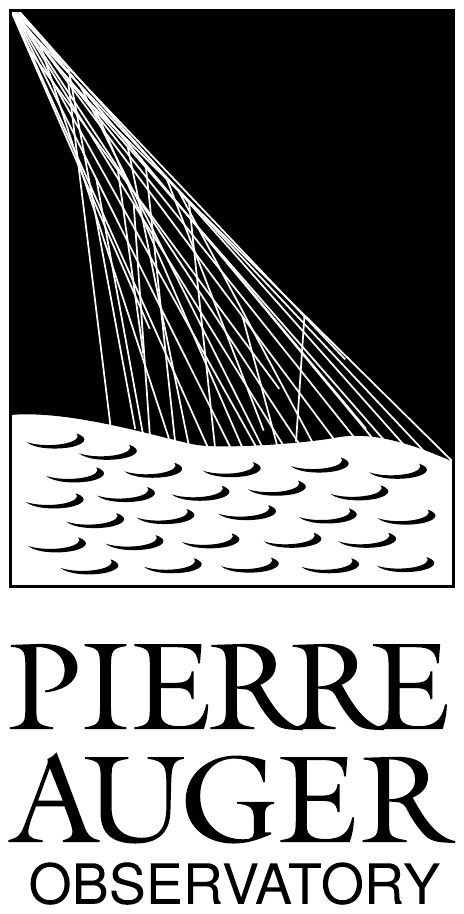}\\[3mm]The Pierre Auger Collaboration}
\affiliation{Av.\ San Mart\'{\i}n Norte 306, 5613 Malarg\"ue, Mendoza, Argentina}

\emailAdd{auger\_spokespersons@fnal.gov}

\abstract{AMIGA (Auger Muons and Infill for the Ground Array) is an upgrade of the Pierre Auger Observatory to complement the study of ultra-high-energy cosmic rays (UHECR) by measuring the muon content of extensive air showers (EAS). It consists of an array of 61 water Cherenkov detectors on a denser spacing in combination with underground scintillation detectors used for muon density measurement. Each detector is composed of three scintillation modules, with 10\,m$^2$ detection area per module, buried at 2.3\,m depth, resulting in a total detection area of 30\,m$^2$. Silicon photomultiplier sensors (SiPM) measure the amount of scintillation light generated by charged particles traversing the modules. In this paper, the design of the front-end electronics to process the signals of those SiPMs and test results from the laboratory and from the Pierre Auger Observatory are described. Compared to our previous prototype, the new electronics shows a higher performance, higher efficiency and lower power consumption, and it has a new acquisition system with increased dynamic range that allows measurements closer to the shower core. The new acquisition system is based on the measurement of the total charge signal that the muonic component of the cosmic ray shower generates in the detector.}


\arxivnumber{2011.06633}

\dedicated{\flushleft\textnormal{\textsc{%
Published in JINST as \href{https://doi.org/10.1088/1748-0221/16/01/P01026}{DOI:10.1088/1748-0221/16/01/P01026}
\\
Report number: FERMILAB-PUB-20-605-AD-E-TD
}}}

\begin{document}
%\linenumbers
\maketitle
\flushbottom

\section{Introduction}
\label{sec:Intro}
The Pierre Auger Observatory~\cite{ThePierreAuger:2015} is located in the North of the city of Malarg\"ue, province of Mendoza, Argentina and covers an area of about 3000\,km$^2$. It is designed to detect ultra-high energy cosmic ray (UHECR) showers with a hybrid detection technique. It consists of 1660 water Cherenkov detectors (WCD)~\cite{Allekotte:2008} arranged on a triangular grid with a distance of $\sim$1.5\,km between stations, conforming the surface detector (SD). It is complemented by 27 fluorescence telescopes (FD)~\cite{ThePierreAuger:2010} located in four buildings on the perimeter of the array observing the atmosphere above the SD. The Pierre Auger Observatory is currently being upgraded~\cite{PAO_DR_up_2016}, and AMIGA~\cite{Etchegoyen:2007,ThePierreAuger:2009a,ThePierreAuger:2009b, Sanchez:2019KN} (Auger Muons and Infill for the Ground Array) is one of the main enhancements. The two main objectives of AMIGA are the measurement of composition-sensitive observables of extensive air showers (EAS) and the study of features of hadronic interactions. That these goals can be achieved with a muon detection techniques was proven by several experiments like KASCADE~\cite{KASCADE:2003} and KASCADE-Grande~\cite{KASCADE-Grande:2010} with big impact on cosmic ray physics.

AMIGA detectors are collocated with 61 WCD, which are deployed on a smaller 750\,m triangular grid in an infilled array of 23.5\,km$^2$ size, conforming the Underground Muon Detector (UMD). Three AMIGA modules of 10\,m$^2$ form an UMD station associated at every WCD. When a cosmic ray shower triggers a WCD, the associated trigger is forwarded to the corresponding UMD station. Each module is buried 2.3\,m below ground to shield the electromagnetic component of cosmic ray showers (vertical shielding of 540\,g/cm$^2$). Every module consists of 64 scintillation bars (of 400\,cm length, 4\,cm width and 1\,cm thickness) and 64 wavelength-shifting (WLS) optical fibers (BCF-99-29AMC from Saint Gobain) of 1.2\,mm diameter, which are glued in a lengthwise groove on each bar. Light produced in the bars by scintillation is absorbed by the WLS fiber. The excited molecules of the fiber decay and emit photons, that are propagated due to total reflection towards the fiber end for detection in a single channel of a multi-pixel photon detector. The module design is optimized for a high light output and thus, for an efficient counting of muons that impinge on the 10\,m$^2$ area. The mechanical AMIGA module design is fixed and detailed in~\cite{Suarez:2016}. 

Until September 2016, an engineering array (EA) of seven UMD stations were equipped with multi-pixel photomultiplier tubes (MPPMT). For the final production of AMIGA modules, the MPPMTs will be replaced with silicon photomultipliers (SiPM)~\cite{Hampel:2017}. The electronics with MPPMT began to be replaced by ones with SiPM since October 2016. Currently, the seven UMD stations of the EA are equipped  with SiPM electronics. 

This paper explains the design and test of the new front-end electronics for the UMD. The design of the front-end electronics for the traditional method of muon counting~\cite{Hampel:2017,Wainberg:2014} for low densities is described in section~\ref{sec:Design overview}. It is followed by the description of the new acquisition system in section~\ref{sec:Integrator} that allows, by measuring the signal charge, the determination of the muon density. In section~\ref{sec:Calibration and performance tests}, laboratory as well as data from the Observatory are used to show how the new acquisition system performs.

\section{Design overview}
\label{sec:Design overview}
This new design is based on the experience gained with the engineering AMIGA system, but it adds the charge measurement as new feature and it improves other parts. The specifications of the new design are detailed in~\cite{Hampel:2017}; they take into account the experience of the engineering design with MPPMT. The objectives for the improved electronics are:
\begin{enumerate}
\item Reduction of the power consumption below 2\,W. The previous design of the front-end with MPPMT consumes 2.4\,W without the photodetector connected.
\item The SiPM readout must be able to trigger on short light pulses of 25-35\,ns width. 
\item The output of each SiPM channel will be digitized as 0-1 signal. The width of the digital output must be similar to the width of the light pulses of the SiPM.
\item The electronic needs to compensate the temperature dependence of the SiPM parameters.
\item The electronic needs to provide a high gain and a low gain signal of the analog sum of all 64 SiPM pixels. 
\end{enumerate}

\begin{figure}[H]
	\centering
	\includegraphics[scale=0.29]{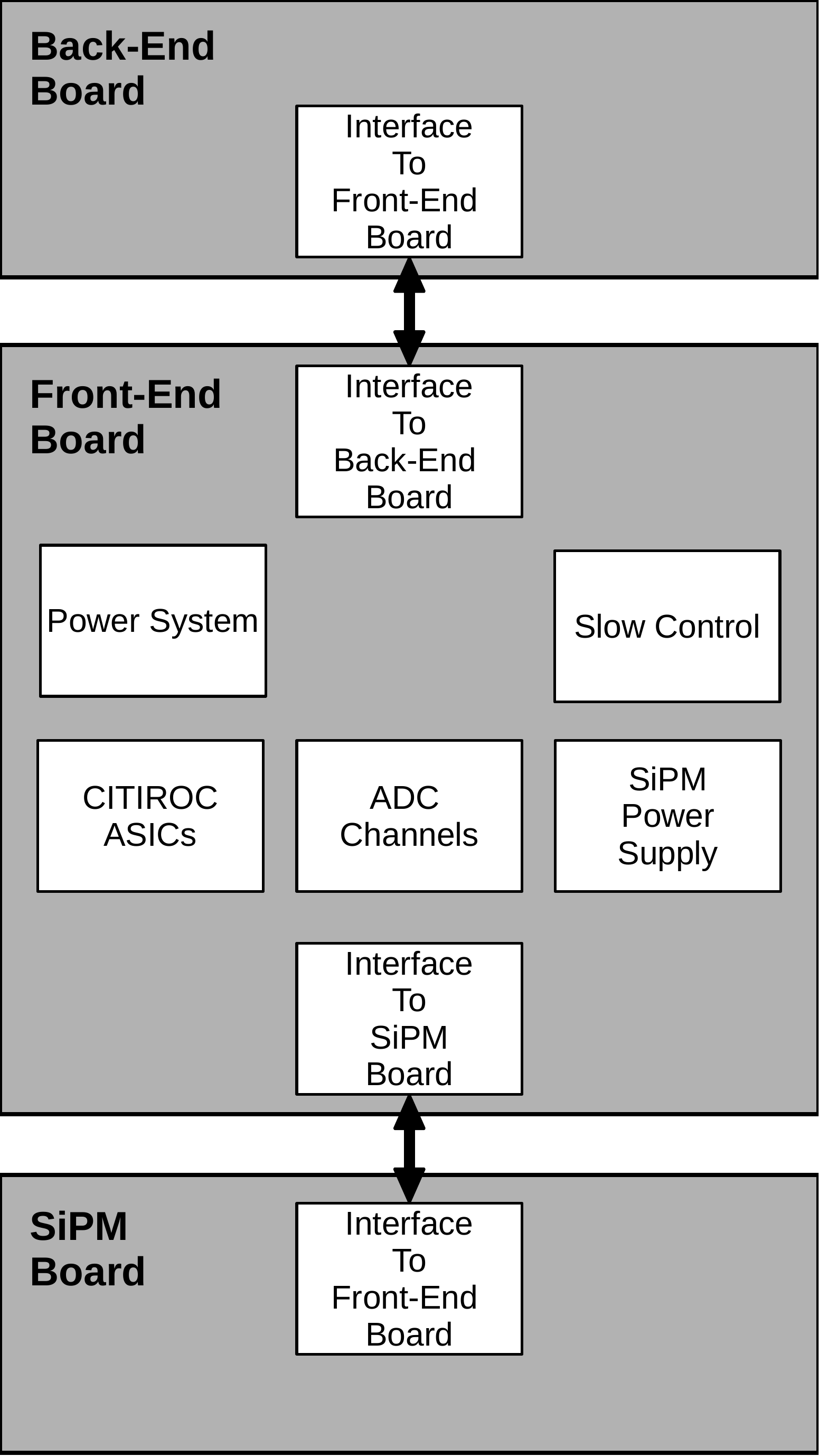}
	\caption{Diagram of the front-end general system and the interfaces to the other boards.}
	\label{fig:general_scheme}
\end{figure}

The diagram of figure~\ref{fig:general_scheme} shows the structure of the new electronics and the interfaces between the different boards.

The electronics is divided in different parts: interface to SiPMs, the SiPM readout, ADC channels (charge measurement), the power supply for the SiPMs (pass through the SiPM interface), the slow control, the power system and the back-end connector (interface to digital electronics). All these parts will be explained in the following sections and subsections. The electronics system of the back-end will be described in a companion future paper. 

\subsection{SiPM readout}
\label{subsec:SiPM_readout}
The SiPM readout is realized by CITIROC\footnote{Cherenkov Imaging Telescope Integrated Read Out Chip.} ASICs\footnote{Application Specific Integrated Circuit.}~\cite{OmegaCITIROC}, special 32-channel integrated circuit designed for the readout of SiPMs~\cite{Hampel:2017}. For the readout of 64 optical channels~\cite{Suarez:2016}, two CITIROC ASICs are required. The chips provide the following functions per channel:
\begin{enumerate}
\item 8-bit digital-to-analog converters (DAC) to adjust the bias voltage ($V_{BIAS}$) of SiPM channels individually.
\item Two pre-amplifier stages with programmable gains for a high-gain and a low-gain output.
\item A fast shaper with 15\,ns peak time per channel. This configuration is recommended for SiPMs by Hamamatsu~\cite{mppc-technote} to decrease the pulse decay time. The smaller pulse width improves the double pulse resolutions and the timing of the trigger electronics.
\item A discriminator following the fast shaper combined with a 1-bit comparator to convert analog signals to digital signals. A 10-bit DAC sets a coarse discriminator threshold common for all 32 channels; it is fine-tuned for each channel by individual 4-bit DACs.
\end{enumerate}

The schematics in figure~\ref{fig:citiroc_sche} (appendix~\ref{appendix_schematics}) shows how the CITIROC chip is interfaced to the electronics. The 32 analog inputs connect directly to the SiPM interface, the 32 trigger outputs to the digital logic. The power system provides a common 5\,V supply, a digital 3.3\,V and an analog 3.3\,V supply. Every supply line is decoupled by several 10\,$\mu$F and 100\,nF capacitors close to the consumer.

\subsection{Power system}
\label{subsec:Power system}
The electronics of the SD and UMD is supplied from a battery-powered system charged through solar panels. Maintenance of the batteries and costs demand  a very low power consumption for the new front end. To meet the design goal of a consumption below 2\,W, a high efficient conversion of the 24\,V battery voltage to several lower consumer voltages is required. Modern switching power regulators provide high conversion efficiencies. However, they are by design prone to spikes and ripples. Linear power regulators on the other hand, provide a high ripple rejection and low noise, but are less efficient at large differences between input and output voltages. 

\begin{figure}[H]
	\centering
	\includegraphics[width=.8\textwidth]{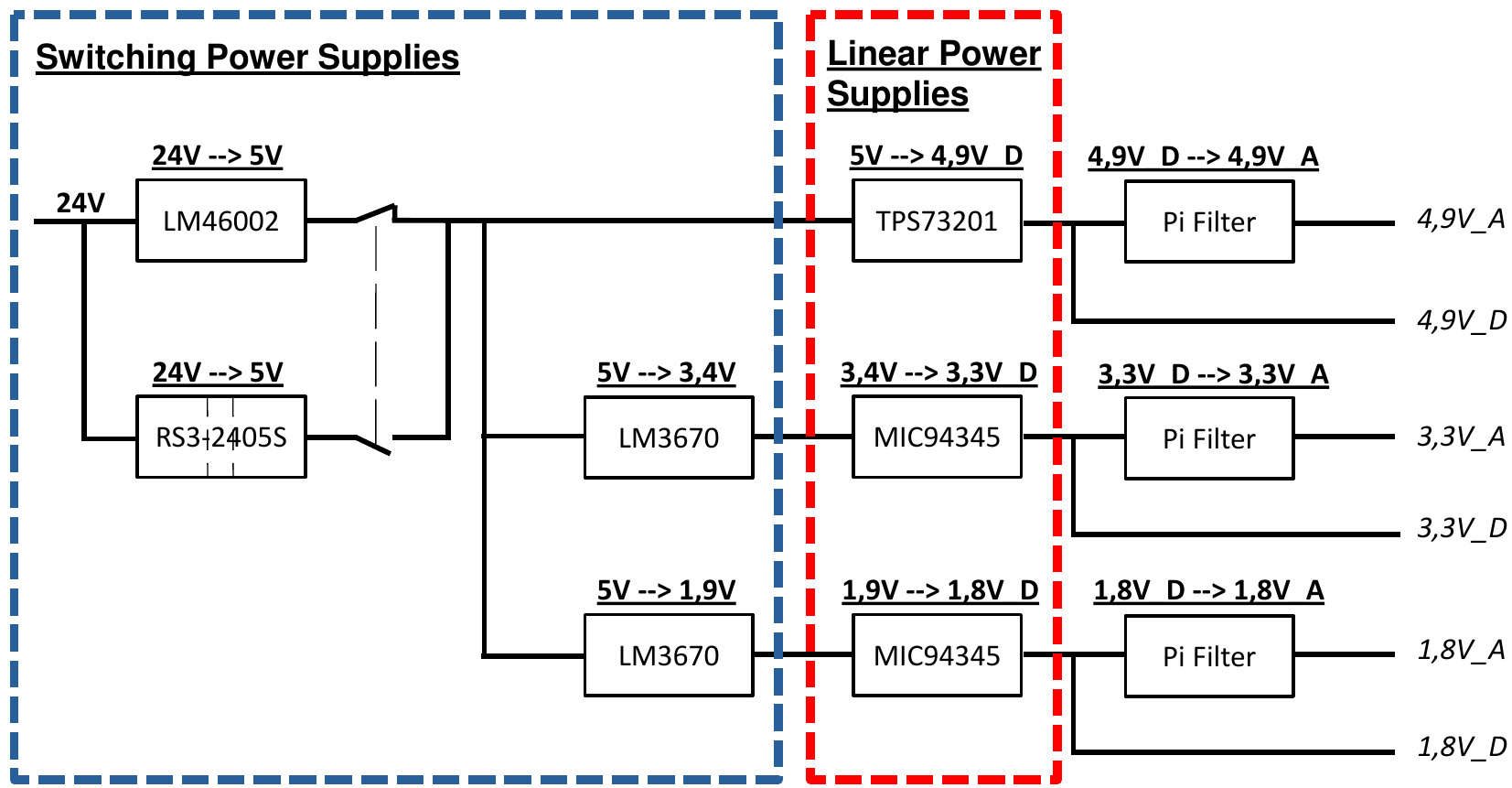}
	\caption{Scheme of power system. In a first stage (dotted blue lines), switching regulators convert the nominal 24\,V input to intermediate voltages (5\,V, 3.4\,V and 1.9\,V). In a second stage (dotted red lines), linear regulators convert to the final voltages (4.9\,V, 3.3\,V and 1.8\,V). The analog voltages are electrically isolated by pi filters from the digital ones.}
	\label{fig:power_scheme}
\end{figure}

The design of the power system (figure~\ref{fig:power_scheme}) combines the advantage of switching regulators and linear regulators in a two stage approach. The first stage made up of switching regulators converts with high efficiency the battery voltage to intermediate voltages of 5\,V, 3.4\,V and 1.9\,V. These intermediate voltages are, in a second stage, converted by linear regulators to the final voltages of 4.9\,V, 3.3\,V and 1.8\,V with low noise.

For the conversion of the nominal 24\,V external battery voltage (23\,V to 32\,V range~\cite{Cancio:2018}) to 5\,V, two different options have been implemented: a non-isolated power supply and an isolated power supply to avoid possible ground loops. The buck controller LM46002~\cite{DS-LM46002} from Texas Instruments is used for the non-isolated power supply, which meets the required specifications of high efficiency ($\eta=88\,\%$ at 150\,mA) and it provides a wide temperature range (-40\textdegree\,C to 125\textdegree\,C). Figure~\ref{fig:non-isolated_supply} (appendix~\ref{appendix_schematics}) shows the schematics for this option.

\begin{figure}[h!]
	\centering
	\includegraphics[width=.7\textwidth]{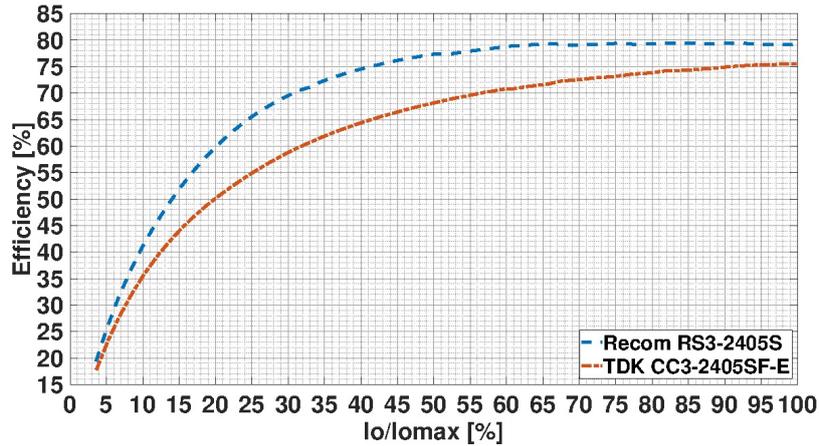}
	\caption{Efficiency measurement of the isolated converters RECOM RS3-2405S (blue line) and TDK CC3-2405SF-E (red line) at an input voltage of 27\,V (mean output voltage of the battery-powered system). The efficiency of the RECOM RS3-2405S is better for all load currents.}
	\label{fig:eff_measurement}
\end{figure}

Two solutions were compared for the isolated power supply option: a) the RS3-2405S~\cite{DS-RS3-2405S} from RECOM and b) the CC3-2405SF-E~\cite{DS-CC3-2405SF-E} from TDK. Both have almost the same properties, including temperature range, input voltage range, maximum output current, low output ripple and isolation voltage. Figure~\ref{fig:eff_measurement} shows the comparison of the converters efficiency for different loads. The converter RS3-2405S has always over the whole current range a higher efficiency compared to the converter CC3-2405SF-E. Therefore, the converter from RECOM was chosen and the schematics is shown in figure~\ref{fig:isolated_supply} (appendix~\ref{appendix_schematics}).

Finally, the power system provides the possibility to choose between the isolated or non-isolated option, but for the better efficiency, lower component count and to avoid possible ground loops, the RS3-2405S is finally selected.

The second part of the switching power supplies are buck controllers of type LM3670~\cite{DS-LM3670} from Texas Instruments. The main characteristics of this controller are low noise, high efficiency and small layout area. The linear regulator stage is delimited by the red dotted lines in figure~\ref{fig:power_scheme}. A low dropout voltage (LDO) and a high ripple rejection are the most important properties of the selected regulators TPS73201~\cite{DS-TPS73201} from Texas Instruments and MIC94345~\cite{DS-MIC94345} from Micrel. The resistors defining the output voltage of the linear regulators are of 0.1\,\% tolerance for high output accuracy and low temperature dependency. 

To avoid introducing noise from digital circuits to the analog part, the analog power supplies are isolated by a pi filter. This also keeps the signal-to-noise-ratio as high as possible and prevents undesirable ringing.

\subsection{Operating temperature range}
\label{subsec:Temperature range}
\begin{figure}[H]
	\centering
	\includegraphics[width=.7\textwidth]{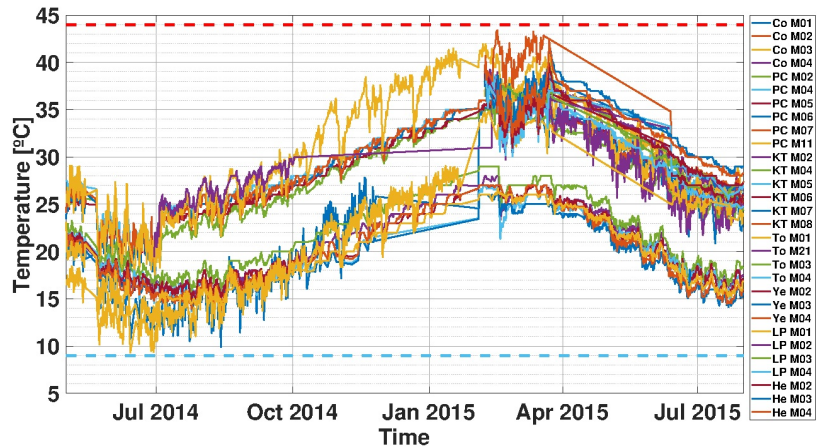}
	\caption{Temperature profile of several AMIGA modules installed in the engineering array. The measurement period from 1st May 2014 to 31st July 2015 covers all seasons (autumn, winter, spring, summer). The minimum temperature value occurs in winter and the maximum value occurs in summer. All values fall within  a temperature band of 9\textdegree\,C (light blue dashed line) to 44\textdegree\,C limits (red dashed line).}
	\label{fig:temp_todos_los_tanques}
\end{figure}

Figure~\ref{fig:temp_todos_los_tanques} shows the temperature profile of all the AMIGA modules of the engineering array as recorded with the back-end electronics in the period May 1st, 2014 to July 31st, 2015. All temperatures fall in the range 9\textdegree\,C and 44\textdegree\,C (dotted blue and red line, respectively). In general, we have selected components of industrial or even military temperature grade for best reliability. However, the SiPM power supply is only available in commercial temperature grade. Nevertheless, the operating temperature range still meets the requirements. 

\subsection{SiPM power supply}
\label{subsec:SiPM_power_supply}
The integrated circuit (IC) C11204-01~\cite{DS-C11204-01} supplies the voltage for the SiPM (type S13081-050CS~\cite{Hampel:2017}) selected for AMIGA. This power supply is recommended by Hamamatsu and allows voltage settings between 50\,V and 90\,V with a resolution of 1.8\,mV. As the SiPM breakdown voltage ($V_{BR}$) varies with temperature, this power supply includes a built-in high precision temperature based voltage compensation system. This compensation system constantly corrects the SiPM operating point in varying temperatures environments, keeping it stable. The operating point is determined by the over-voltage defined as $\Delta V=V_{BIAS}-V_{BR}$. The IC provides also an output voltage and an output current monitor for slow-control. All these functions are controlled from the back-end board via a serial interface (UART). Due to the digital voltage levels of the back-end board (3.3\,V maximum) and the power supply digital voltage levels (5\,V powered), a logic level shifter in each UART signal is needed to adapt the different voltage levels. The power supply provides an overcurrent protection, which switches the output voltage off for output currents above 3\,mA and duration of more than 4\,s. As all 64 SiPM pixels are supplied from a single supply, the inrush current at power-on would trigger this overcurrent protection. To avoid this problem, the limiting resistor of the pi filter (R18 in the schematics figure~\ref{fig:hv_power_supply} of appendix~\ref{appendix_schematics}) is increased from the recommended value 10\,\(\Omega\) to 1\,k\(\Omega\).

\subsection{Back-end Connection}
\label{subsec:Connection to back-end}
For the connection with the back-end, two different types of connectors are used. One is a high-speed connector for the CITIROC signals and ADC channel signals (section~\ref{subsubsec:High speed connector}), the other one is a power connector for the battery voltage and power on/off control signal for the power supplies (section~\ref{subsubsec:Power connector}).

\subsubsection{High speed connector}
\label{subsubsec:High speed connector}
The selected high speed connector QTH-090-04-L-D-A-K-TR~\cite{DS-QTH} from Samtec provides a wide bandwidth of 9\,GHz/18\,Gbps and an inner ground plane for improved noise isolation. CITIROC signals, ADC channel signals, with their respective control signals and I$^{2}$C bus signals (see section~\ref{subsec:Slow control system}) pass through this connector.

\subsubsection{Power connector}
\label{subsubsec:Power connector}
A power connector is used in this design to eliminate connections cables between the frond-end board and the back-end board. The external battery power is connected to the back-end board and from there distributed to the front-end. 
The selected power connector HW-06-20-L-D-630-120~\cite{DS-HW} from Samtec provides high current capability and the same height as the high speed connector, allowing board stacking. As shown in figure~\ref{fig:power_connector} (appendix~\ref{appendix_schematics}), the battery voltage and the on/off control signal for the power supplies pass through this connector. To avoid possible ground loops between the two boards and have separate grounds, an isolated solid stage relay is used for the power supply control stage (CTRL1\_PS and CTRL2\_PS).

\subsection{Interface to SiPM}
\label{subsec:Connection to SiPM board}
The connector QMSS-026-06.75-LD-PT4~\cite{DS-QMSS-PC} from Samtec is the interface to the SiPM. Due to the low amplitude of the analog signals from the SiPM, special care has to be taken to avoid introducing noise. This connector is a shielded ground plane header with a large bandwidth of 6\,GHz/12\,Gbps. It integrates eight pins for the supply voltages and ground connection.

\subsection{Slow control system}
\label{subsec:Slow control system}
The analog to digital converter (ADC) ADS7828~\cite{DS-ADS7828} from Texas Instruments is the heart of the slow control system. It supervises all on board (analog and digital) supply voltages and the temperature of the CITIROC ICs as shown in figure~\ref{fig:slow_monitor} (appendix~\ref{appendix_schematics}).
The ADS7828 is a single-supply, low-power, 12-bit data acquisition device with a 50\,kHz sampling frequency that features a serial I$^{2}$C interface and an 8-channel multiplexer.
An external voltage reference of 2.048\,V, instead of the 2.5\,V internal reference, provides a stable ADC resolution of 0.5\,mV. The SiPM variables, such as the power supply output voltage, the power supply output current and the SiPM temperature, are monitored through the Hamamatsu power supply chip, and are retrieved via an UART interface. 

\subsection{Layout}
\label{subsec:Layout}
As this design is a mixed signal system, special care had to be taken in the printed circuit board (PCB) layout to avoid introducing digital noise into the analog stage. As the SiPM signals are of very low amplitude (around 8\,mV typically), a very good signal to noise ratio (SNR) is required. The disposition strategy of the ground planes, explained in~\ref{subsubsec:Ground planes}, and the controlled impedance transmission lines, explained in ~\ref{subsubsec:Controlled impedance}, are of special importance. 

\subsubsection{Ground planes}
\label{subsubsec:Ground planes}
\begin{figure}[h]
	\centering
	\includegraphics[width=0.6\textwidth]{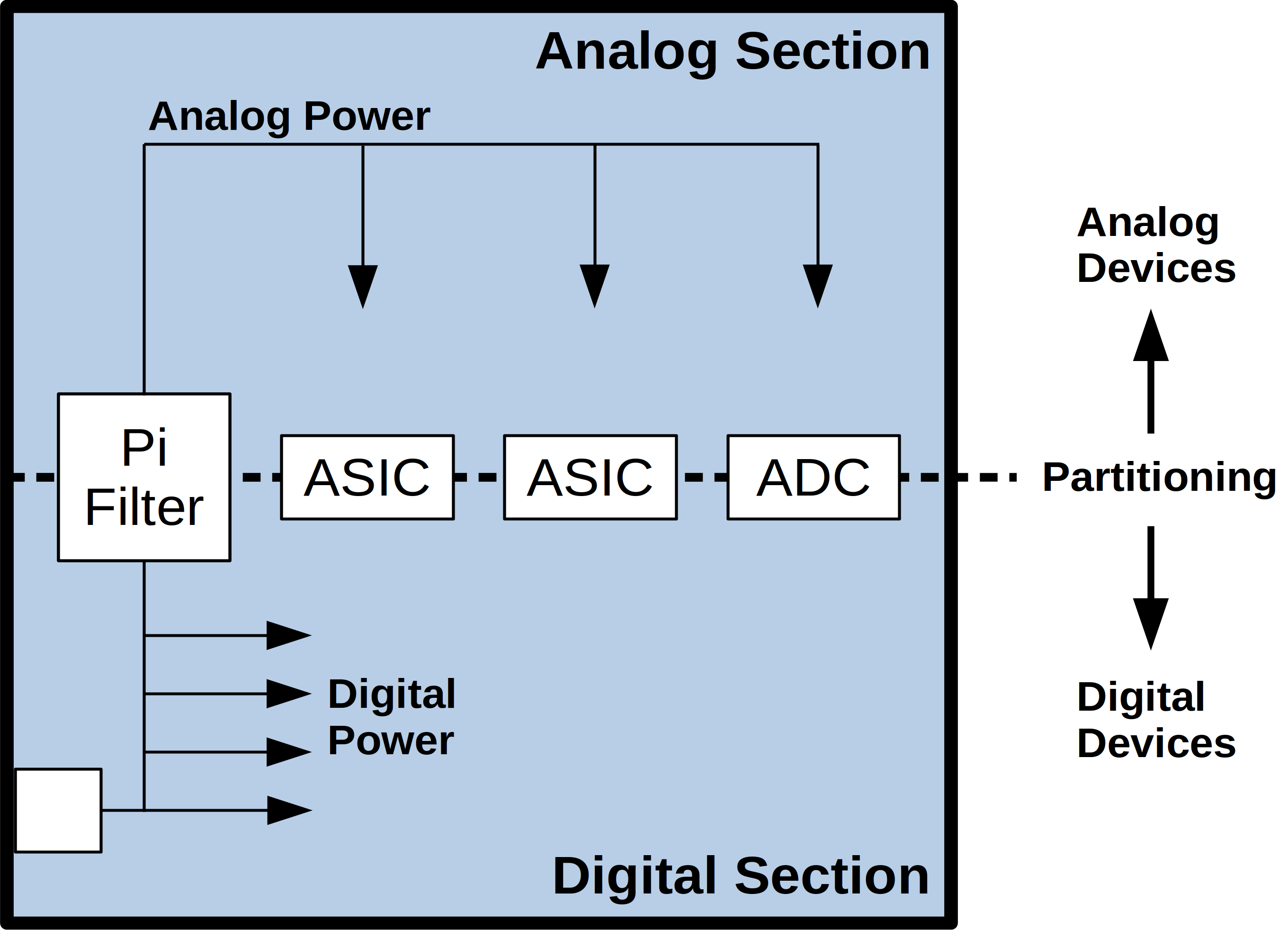}
	\caption{Power and ground strategy for PCBs with multiple ASICs, ADC and digital devices. There is only one ground plane for all the devices. The analog and digital section are separated in different regions. Thus, the return current is restricted to the corresponding section, avoiding noise induction from the digital to the analog section.}
	\label{fig:ground_strategy}
\end{figure}

The layouts of ground and supply planes are based on application notes~\cite{AN-ground_plane_part1,AN-ground_plane_part2} from Texas Instruments, where the technique of partitioned ground planes is suggested for designs with many mixed grounds (analog and digital) devices. This technique uses a single ground plane in the whole PCB instead of having a separate ground plane for analog devices and for digital devices. 
In this way, analog signals are routed only in the board analog section, and digital signals are routed only in the board digital section, both on all layers. Under these conditions, the digital return currents do not flow in the analog section of the ground plane and remain under the digital signal trace, not inducing noise in analog signals. In figure~\ref{fig:ground_strategy} is shown the PCB layout diagram.

\subsubsection{Controlled impedance lines}
\label{subsubsec:Controlled impedance}
As this design includes high speed signals (for example, ADC channel analog signal tracks), lines of controlled impedance are required. The impedance of transmission lines depends on the PCB materials (dielectrics and conductors) and their geometry. As the maximum working frequency is around 200\,MHz, the dielectric FR4 (\(\epsilon_{r}=4.4\)) was chosen as the PCB material, which is recommended for frequencies up to 500\,MHz. Asymmetrical striplines, differential striplines, microstrips and differential microstrips are used. The geometry and characteristic of each of them was designed with the help of tools detailed in \cite{Asym-stripline-cal,Diff-stripline-cal,Microstrip-cal,Diff-microstrip-cal}. Single-ended signal tracks have a characteristic impedance of 50\,\(\Omega\) and differential signal tracks have a characteristic impedance of 100\,\(\Omega\). A 10-layer FR4 PCB with a thickness of 93\,mil (2.36\,mm) ensures the calculated geometries and the impedance values of the transmission lines.

\section{New acquisition system: charge measurement} 
\label{sec:Integrator}
\begin{figure}[h!]
	\centering
	\includegraphics[width=0.6\textwidth]{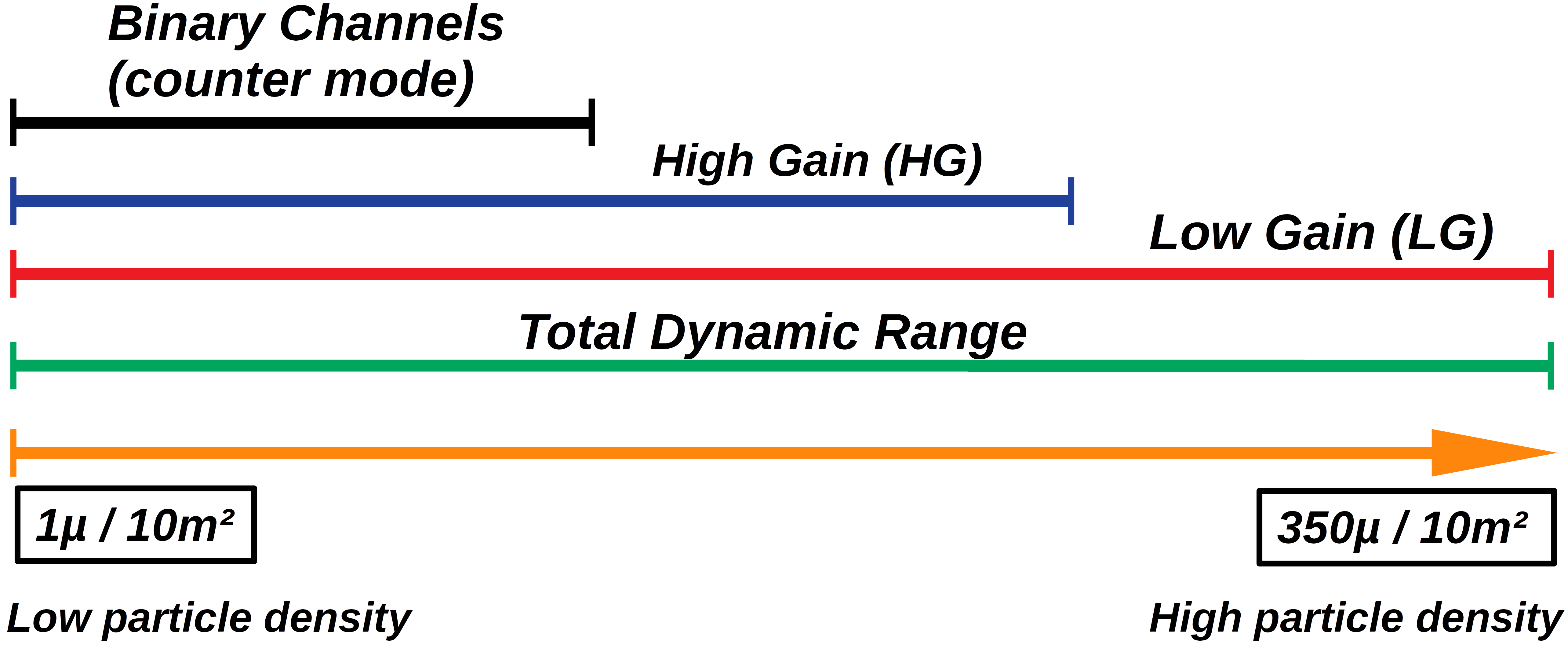}
	\caption{A greatly extended particle density range is achieved by overlapping operating ranges of the binary channels and the LG and HG channels.}
	\label{fig:total_dynamic_range}
\end{figure}

The measurement of the total charge is an enhancement of the AMIGA front-end to extend its dynamic range for particle detection. Thus, each module that make up the UMD will be able to measure in regions of high particle density. These regions are located close to the impact point of the core of the air shower. The new acquisition system determines the total charge generated by the muonic component of the cosmic ray showers during its impact on the UMD. It is implemented in two independent channels of different resolution: a high gain channel (HG) and a low gain channel (LG). Figure~\ref{fig:total_dynamic_range} shows the operating ranges of HG channel, LG channel and the binary channels (counter mode). 

The operating ranges of the HG and LG start at the same low particle density value as the binary channels~\cite{Wundheiler:2020}, but reach higher values than these. Simultaneously, the LG channel saturates for higher signal charges than the HG channel, allowing to measure up to very high particle densities. That the three ranges start at the same particle density value allows firstly, to obtain complementary measurements at those values and secondly, to perform the comparison of the binary channels with the HG and LG channels. The main difference in the overlapping ranges is the resolution. The binary channels perform with better resolution at very low particle densities, while the resolution of the HG and LG channels improves as the number of particles arriving at the detector increases. As a result, the measurements range of the AMIGA detector is extended considerably with respect to the previous design implemented for the engineering array. The binary channels saturate when all 64 scintillators have simultaneous signals but, as it will be described in the following sections, the HG and LG channels might measure up to 85 and 362 simultaneous muons without losing its linearity, respectively.

\subsection{HG and LG setup (ADC channels)}
\label{subsec:Scheme overview}
\begin{figure}[h]
	\centering
	\includegraphics[width=0.85\textwidth]{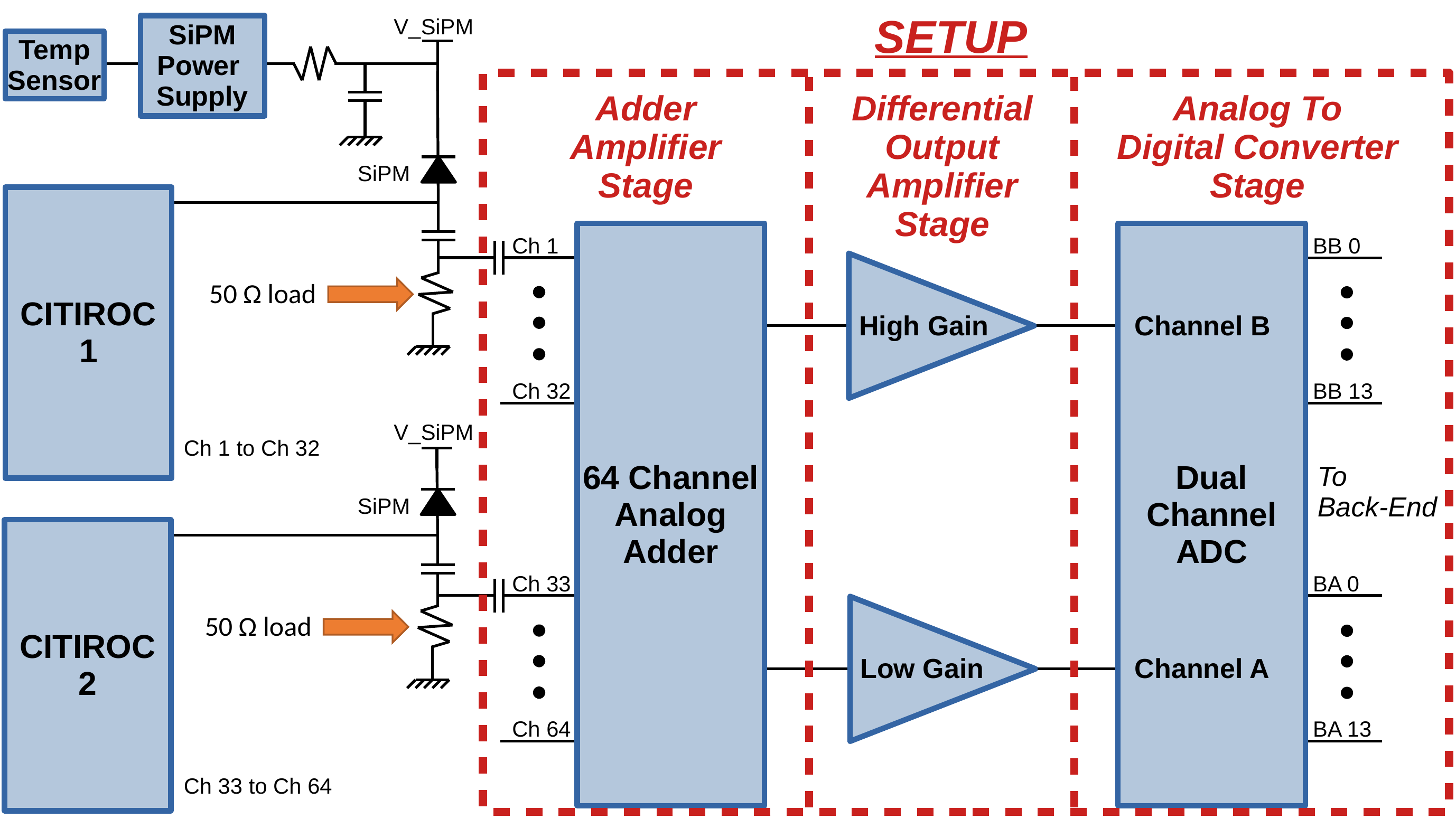}
	\caption{Scheme of the HG and LG channels. The signals from the 64-pixel SiPM array are added in a first stage (Adder Amplifier Stage), amplified and splitted in a high gain/low gain branch (Differential Output Amplifier Stage) and continuously digitized by a dual ADC (Analog To Digital Converter Stage). This circuitry represents a load of 50\,$\Omega$ to the CITIROC inputs.}
	\label{fig:integrator_scheme}
\end{figure}

The determination of the total charge is realized into three stages: the Adder Amplifier Stage, the Differential Output Amplifier Stage and the Analog To Digital Converter Stage as shown in figure~\ref{fig:integrator_scheme}.

The adder amplifiers built up the analog sum of the 64 SiPM channels. The signals are tapped at the CITIROCs inputs providing a load of 50\,$\Omega$ without affecting the CITIROCs functions. The summing of all 64 channels is realized by 4 amplifiers, which adds up groups of 16 signals very close to the two SiPM connectors (two amplifiers per connector). In this way, mismatches and unwanted signal reflections are avoided. The resulting sum-signals are again summed into a last amplifier which is located near the second stage. In this second part, a calibration channel is also added with a SMB connector that connects to this last amplifier. The schematics of the first stage is shown in figure~\ref{fig:adder_amplifiers} (appendix~\ref{appendix_schematics}). The five adders are realized by the dual amplifier AD8012~\cite{DS-AD8012} from Analog Devices. A positive DC voltage at the positive input shifts the amplifiers operating point such that a dual supply voltages can be avoided.

The following Differential Output Amplifier Stage splits the result of the summation of the 64 SiPMs signals into two independent channels: the low (LG) and high (HG) gain. Differential output amplifiers are used to have a better immunity to noise and to take advantage of the ADC features of the next stage. Like in the previous stage, a negative supply voltage is avoided by a shift of the operation point to the mid value of the power supply of 4.9\,V (2.45\,V). The used amplifier is the LMH6550~\cite{DS-LMH6550} from Texas Instruments. The schematics of this stage with the two channels are shown in figure~\ref{fig:opamp_diff_output} (appendix~\ref{appendix_schematics}).

The Analog To Digital Stage is the last one. The dual channel ADC ADS4246~\cite{DS-ADS4246} from Texas Instruments digitizes the differential LG and HG signal with 160\,Msps, thereby consuming only 332\,mW total power. This sampling rate is half of the SiPM back-end sample rate for the binary channels (320\,MHz). The input low pass filter is designed for a cutoff frequency of 80\,MHz to avoid aliasing. The input driver circuit used is the one recommended by the datasheet to reduce glitches and noise, named~\emph{"Drive Circuit With Low Bandwidth (For Low Input Frequencies Less Than 150\,MHz)"}. All the output bit and status pins of the two channels are protected from large capacitive load currents by 49.9\,$\Omega$ series resistors. The schematics of this stage is shown in figure~\ref{fig:adc_sche} (appendix~\ref{appendix_schematics}). 

\subsection{Simulation}
\label{subsec:Simulation}
\begin{figure}[H]
\begin{subfigure}{.5\textwidth}
  \centering
    \includegraphics[width=\linewidth]{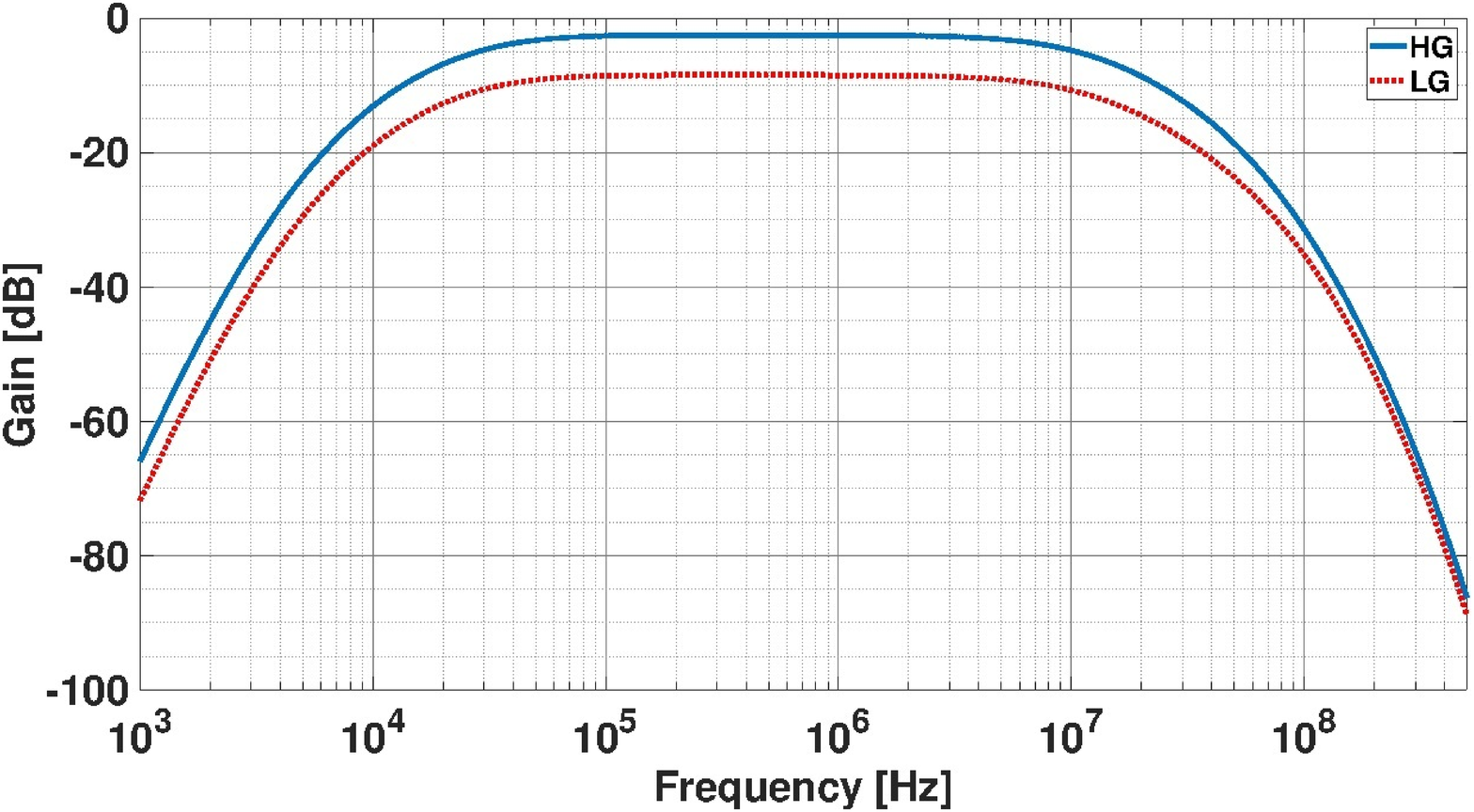}
\end{subfigure}
\begin{subfigure}{.5\textwidth}
  \centering
    \includegraphics[width=\linewidth]{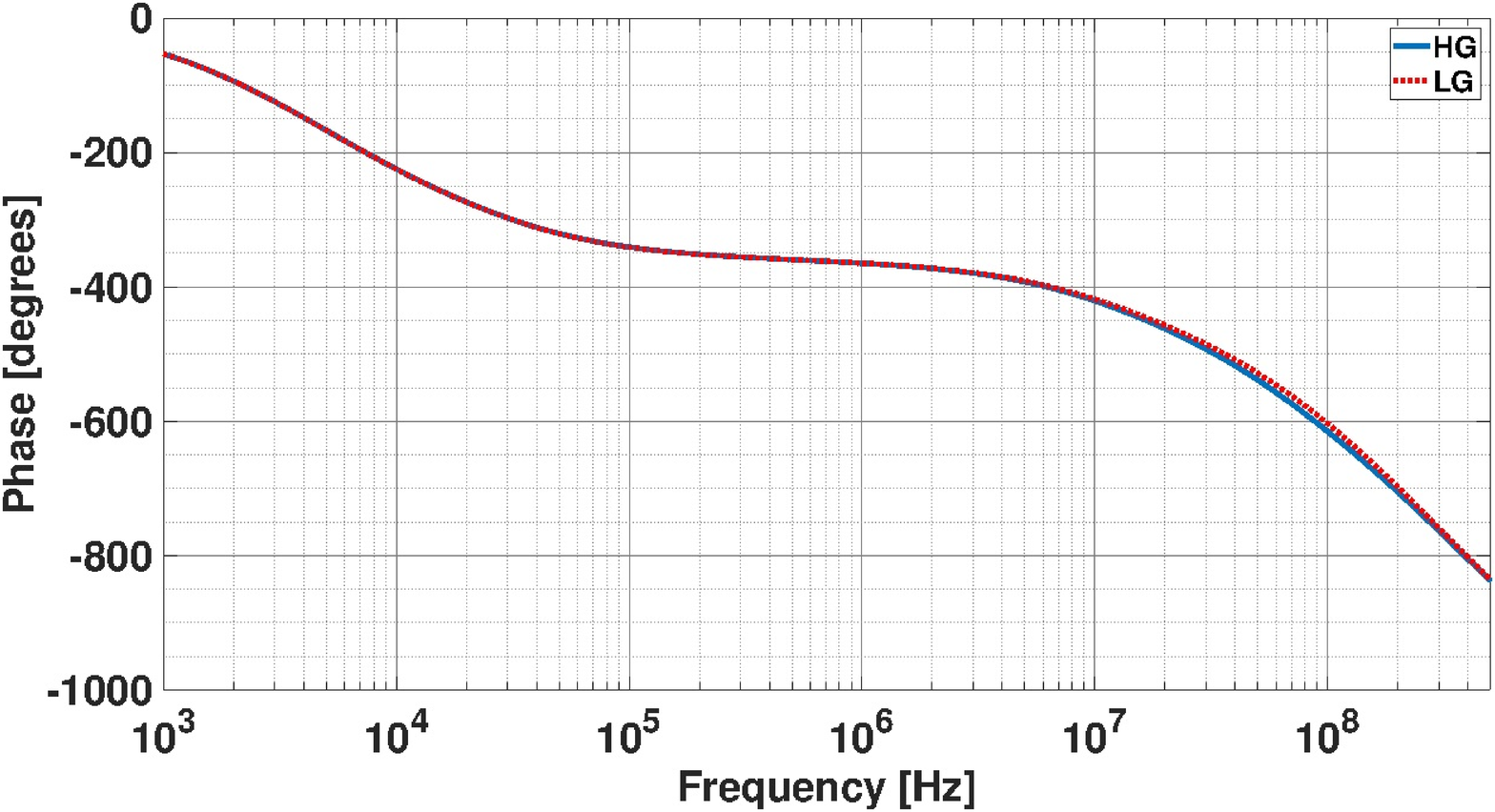}
\end{subfigure}
  \caption{Bode diagram of HG and LG channel. Both channels have similar frequency response, only differing in the gain in the pass band by 6\,dB. The bandwidth is about 11.80\,MHz for both channels.}
  \label{fig:transfer_hg_lg_all}
\end{figure}

Every stage of the HG and the LG channels was modelled with the open-source circuit simulator program SPICE~\cite{Nagel:1973,Nagel:1975,Quarles:1989}. The simulation provided the Bode diagrams with gain and phase for every stage. A transfer function was fitted to every diagram and the overall frequency response for the full chain of HG and LG was derived as shown in figure~\ref{fig:transfer_hg_lg_all}. The 3\,dB bandwidth determined from the plot is 11.80\,MHz for both channels (from 25\,kHz to 11,87\,MHz). The maximum pass band gain is -2.50\,dB for HG and -8.47\,dB for LG, respectively. Both channels have the same phase response. 

\subsection{SiPM front-end printed circuit board}
\label{subsec:pcb}
\begin{figure}[H]
	\centering
	\includegraphics[width=\textwidth]{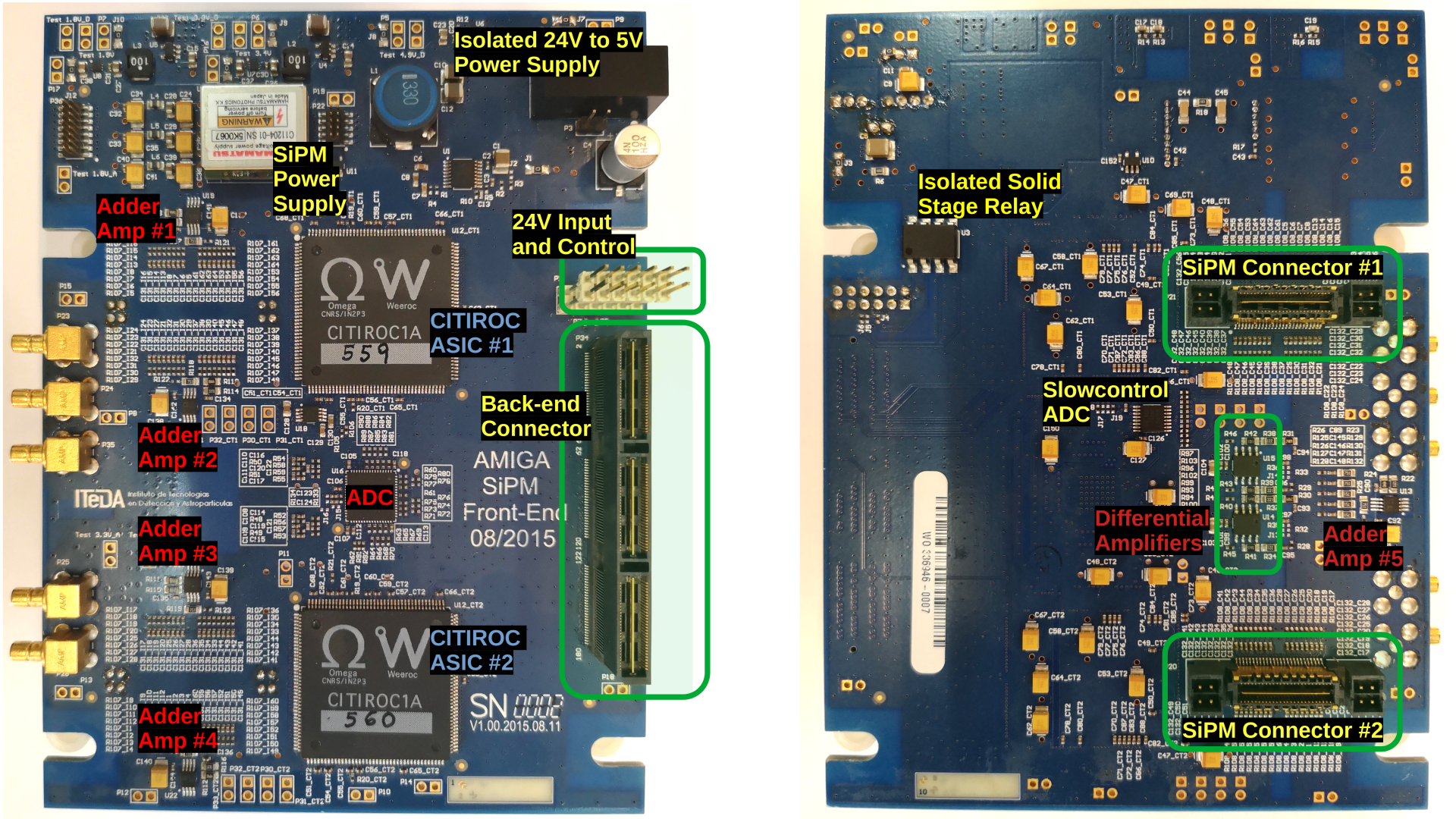}
	\caption{AMIGA SiPM Front-end board, showing the main components and main connectors of the top (left) and bottom (right) side. The components of the charge acquisition system are marked in red.}
	\label{fig:pcb}
\end{figure}

The pictures of the front-end board are shown in figure~\ref{fig:pcb}. This figure shows the layout of the board and indicates the locations of the main connectors and components.
 
\section{Laboratory tests of the ADC channels}
\label{sec:Calibration and performance tests}
To verify the design of the ADC channels (HG and LG) and measure its performance, two different test setups have been built up. The first one was performed with signals coming from atmospheric muons penetrating scintillator bars. How the mean charge is obtained from the muon signal with this setup is described in section~\ref{subsec:detector_calibration_setup}. The second test - based on an array of 64 LEDs - is described in section~\ref{subsec:Linearity_characterization_setup}. The charge calculation algorithm is explained in section~\ref{subsec:Charge calculation algorithm}. 

\subsection{Charge calculation algorithm}
\label{subsec:Charge calculation algorithm}
\begin{figure}[h]
\begin{subfigure}{.48\textwidth}
  \centering
  \includegraphics[width=\linewidth]{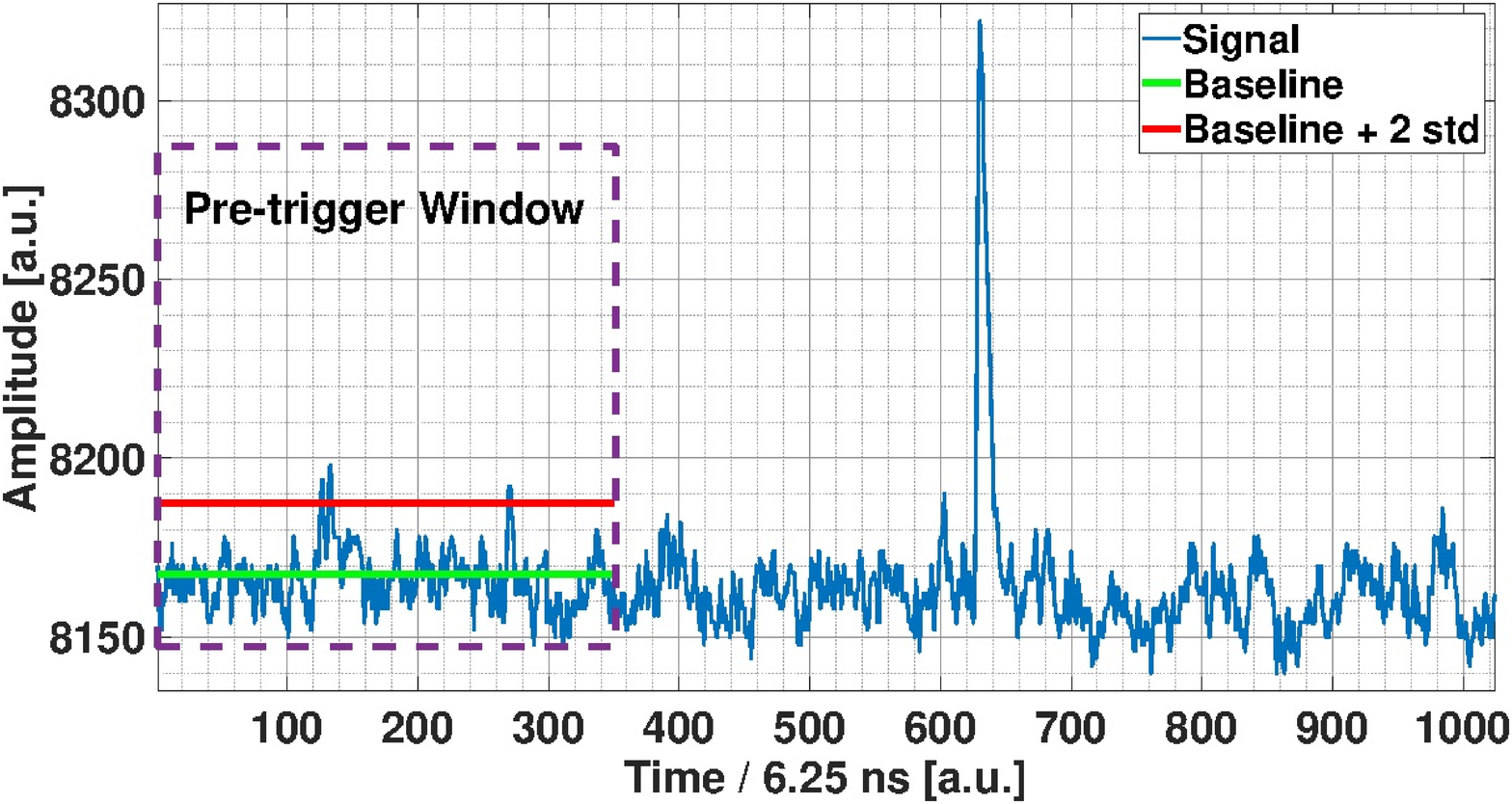}
  \caption{HG channel.}
  \label{sfig:hg_example_pulse}
\end{subfigure}
\begin{subfigure}{.48\textwidth}
  \centering
  \includegraphics[width=\linewidth]{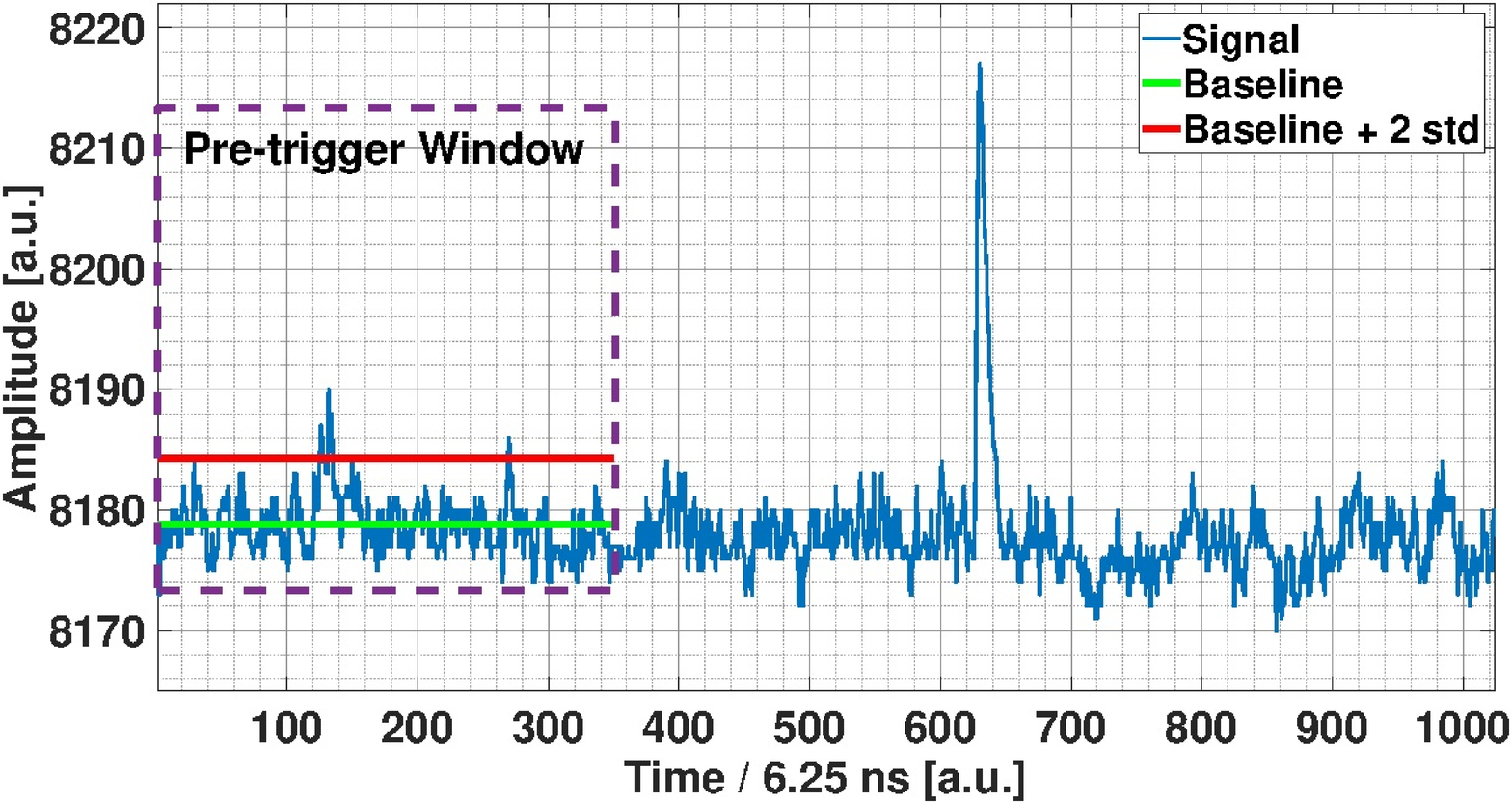}
  \caption{LG channel.}
  \label{sfig:lg_example_pulse}
\end{subfigure}
\caption{Typical traces of the HG and LG channels. The total number of samples of each trace is 1024.}
\label{fig:integrator_example_pulse}
\end{figure}

Typical signal traces of the HG and LG channels are shown in figure~\ref{fig:integrator_example_pulse}. Each trace consists of 1024 samples with a 6.25\,ns separation (corresponding to 160\,MHz sampling rate). The recorded traces are triggered at around sample \# 600 where the pulses start. In a first step, the baseline B and the standard deviation $\sigma$ are derived from the first 350 samples in the pre-trigger window. The signal charge is then calculated in the post-trigger window from the raw data corrected by the baseline B determined in step 1. All baseline corrected values are added up for those bins where the signal is more than 2 $\sigma$ away from baseline. The value determined this way is regarded as the signal charge, the quantity considered in the rest of the chapter.   
  
\subsection{Detector characterization setup}
\label{subsec:detector_calibration_setup}
\begin{figure}[h]
	\centering
	\includegraphics[width=0.9\textwidth]{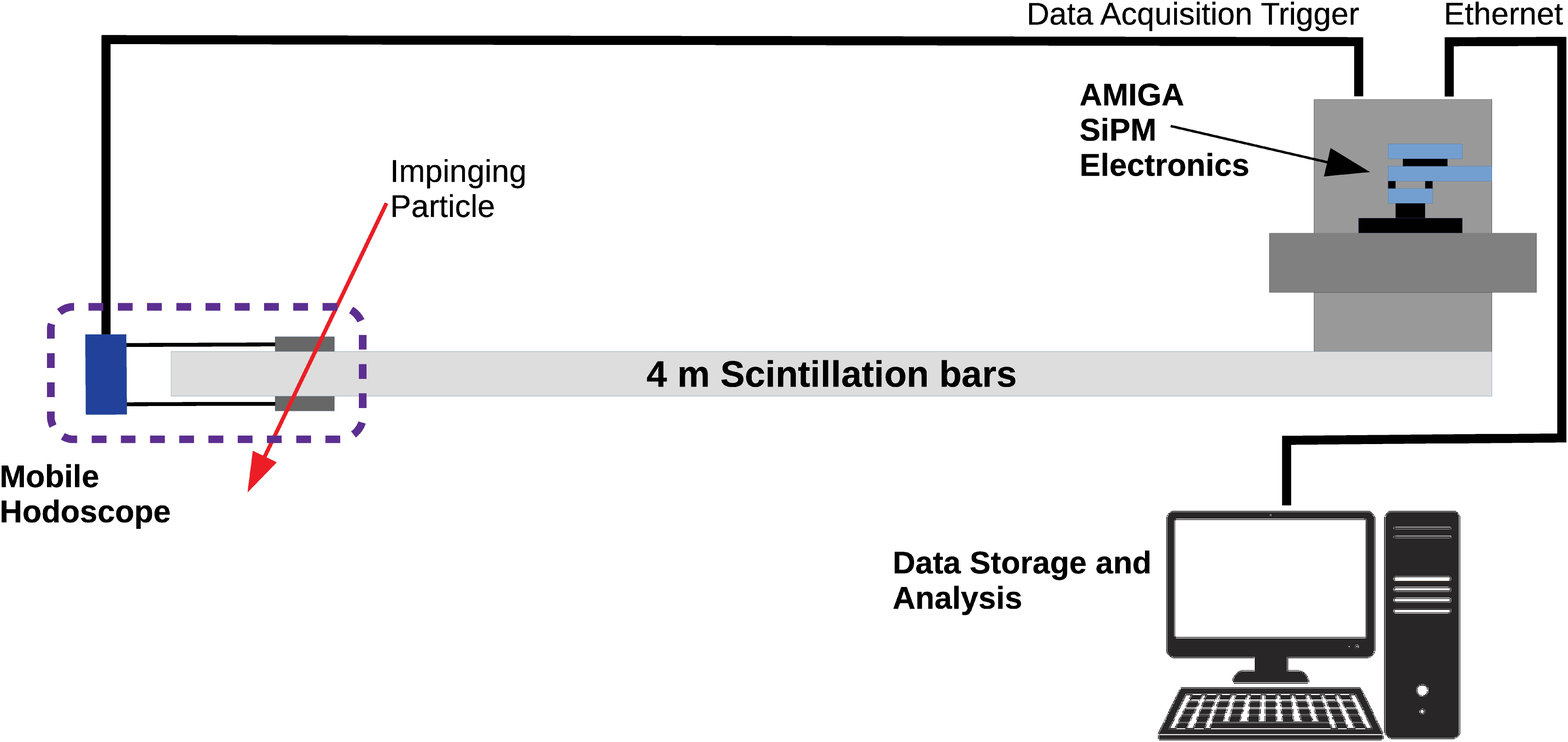}
	\caption{Scheme of setup for muon characterization: Particles penetrating the mobile hodoscope generate the data acquisition trigger for the AMIGA electronics. The readout PC is responsible for the data storage and processing.}
	\label{fig:pipa4_scheme}
\end{figure}

The main components of the setup for the detector characterization are shown in figure~\ref{fig:pipa4_scheme}. The full AMIGA electronics with SiPM board, the new front-end board and the back-end board form the device to be calibrated. The parameters of the electronics like SiPM supply voltage and the configuration of the CITIROC ASICs were obtained in advance by the calibration method described in~\cite{Hampel:2017}. The SiPM electronics provides the ability to turn individual channels on or off as needed for a specific measurement. A mobile hodoscope~\cite{Platino:2011} (at the left in figure~\ref{fig:pipa4_scheme}) provides a trigger, whenever a particle penetrated the device and the main scintillator bar. It consists of two scintillator sheets of dimensions 4\,cm x 4\,cm x 1\,cm coupled to a SiPM for light detection. Both sheets are aligned with a separation distance of 5 cm and the main scintillation bars placed in between. This configuration selects a particular solid angle of muons penetrating the main scintillators.

The trigger condition is fulfilled when the upper and lower SiPM signals from the hodoscope exceed a programmable threshold within a coincidence time of 60\,ns. When this happens, the generated trigger enters the AMIGA electronics and initiates the event acquisition and storage in the internal RAM memory. Finally, the events are readout by a server PC, where all the AMIGA programs (acquisition and monitoring systems) are running. The stored events are analysed and processed offline by this server. 

The scintillation system consists of eight bars stacked in a configuration of two columns of four bars each. For our tests, we have used only the four bars of one column. They were made of identical material like the final AMIGA production module for best reproduction of the situation at the Pierre Auger Observatory site. Those bars are coupled by optical fibers (same as AMIGA) to four non-adjacent pixels of the SiPM to avoid crosstalk effects.

\subsubsection{Muon mean charge}
\label{muon_events_charge}
\begin{figure}[h!]
  %\centering
  \begin{subfigure}{.5\linewidth}
    \centering
    \includegraphics[width=\linewidth]{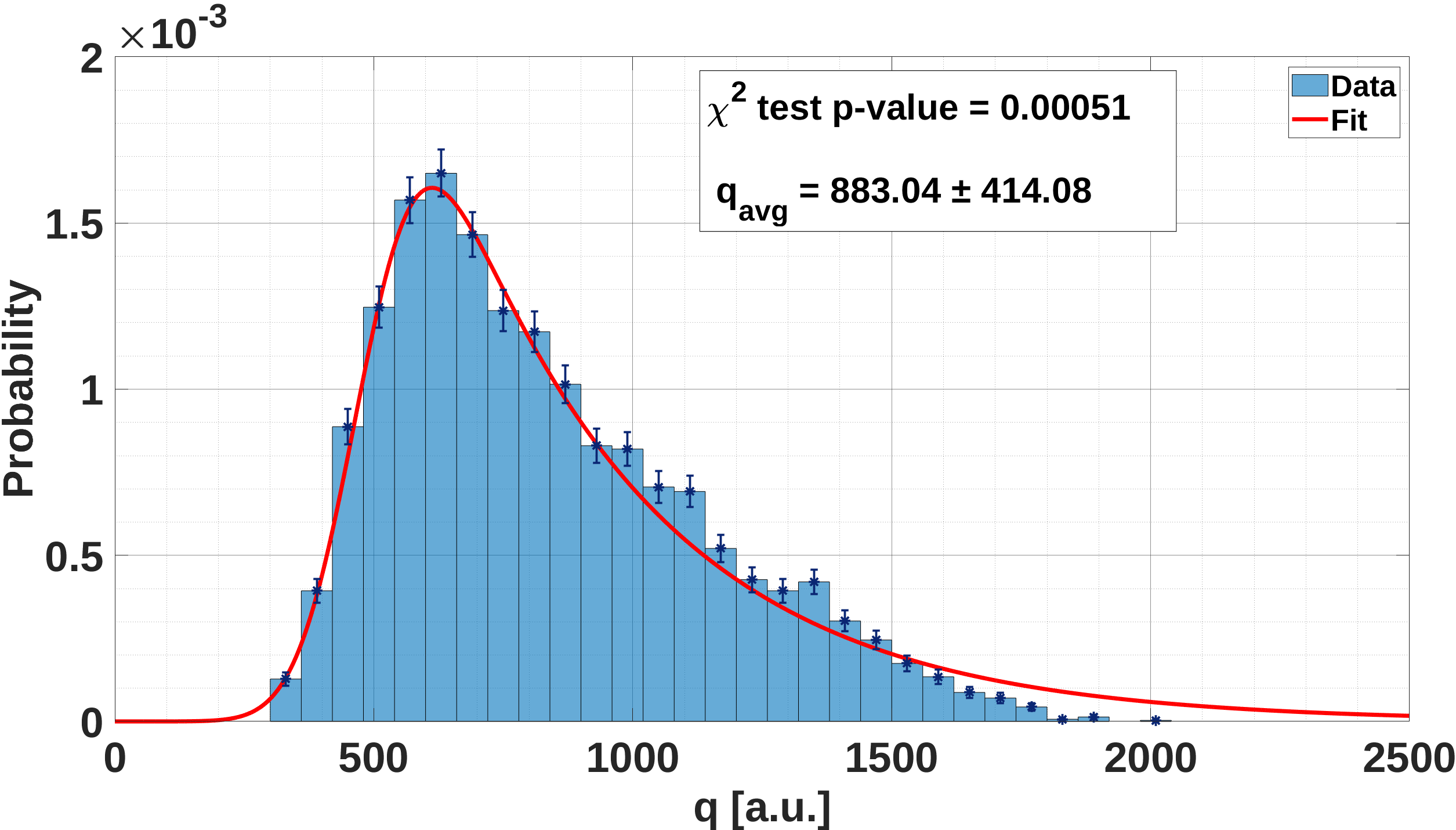}
    \caption{HG channel.}
  \end{subfigure}
  \begin{subfigure}{.5\linewidth}
    \centering
    \includegraphics[width=\linewidth]{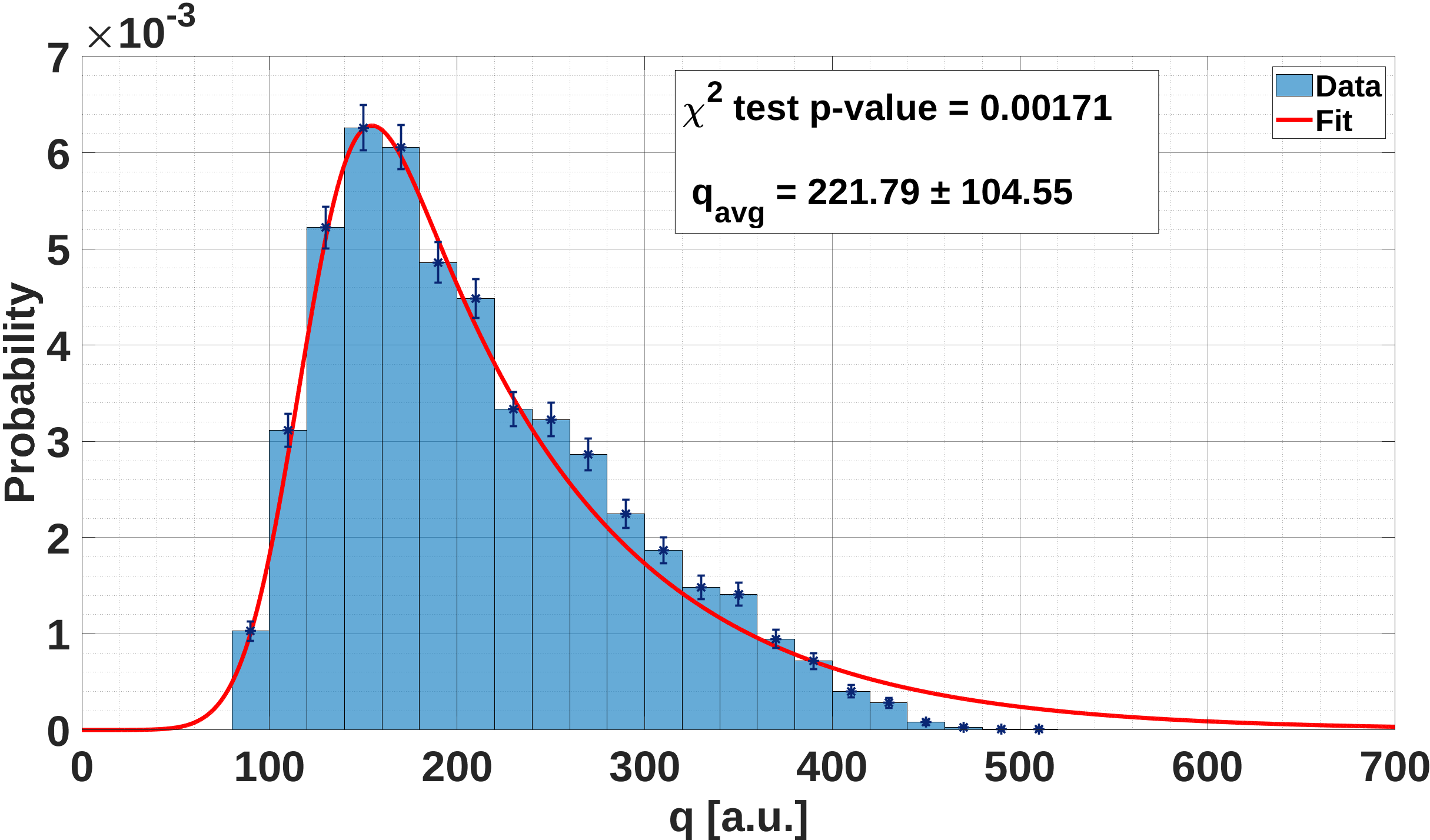}
    \caption{LG channel}
  \end{subfigure}  
  \caption{Muon charge histograms. The mean muon charge $q_{avg}$ is derived from the fit.}
  \label{fig:muon_charge_setup_1} 
\end{figure}

Due to the light attenuation in the fiber, the signal amplitude at the SiPM depends on the position where the muons penetrate the bars. In a pre-analysis, we have averaged the pulse charge over several events and several positions along the bar (at 1\,m, 1.5\,m, 2\,m, 2.5\,m, 3\,m, 3.5\,m, 4\,m and 4.5\,m of optical fiber distance to the SiPM pixels). The measurement was performed with four enabled pixels corresponding to the four bars in the column. The charge measured at the ADC channels corresponds to a charged particle crossing the four scintillator bars, the total charge seen by the ADC channels is divided by four. Figure~\ref{fig:muon_charge_setup_1} shows the charge distribution for the HG channel (a) and LG channel (b). The distribution contains events from several positions along the bar measured for the same duration. $q_{avg}$ is the mean value of the fitted distribution.

Every histogram was fitted with the exponentially modified Gaussian (EMG) distribution (exGaussian distribution) whose probability density function is given by equation~(\ref{eq:exgauss}). This density function is a convolution of normal and exponential probability density functions.

\begin{equation}
f(x;\mu,\sigma,\lambda)=\frac{\lambda}{2} \, e^{\frac{\lambda}{2}(2\mu+\lambda\sigma^2-2x)} \, \textrm{erfc}(\frac{\mu+\lambda\sigma^2-x}{\sqrt{2}\sigma})
\label{eq:exgauss}
\end{equation}

Here \textit{x} is the random variable, $\mu$ and $\sigma$ are the mean and the standard deviation of the Gaussian distribution, $\lambda$ the attenuation parameter of the exponential distribution and \textit{erfc} is the complementary error function defined by equation~(\ref{eq:erfc}).  

\begin{equation}
\textrm{erfc}(x)=\frac{2}{\sqrt{\pi}}\int_x^\infty e^{-t^2}dt
\label{eq:erfc}
\end{equation}

The value $q_{avg_n}$ is the mean of the fitted exGaussian distribution. The mean and the standard deviation of the distribution are obtained by equations~(\ref{eq:exgauss_mean}) and (\ref{eq:exgauss_std}), respectively. 
\begin{equation}
\textrm{MEAN}=q_{avg}=\mu+\frac{1}{\lambda}
\label{eq:exgauss_mean}
\end{equation}

\begin{equation}
\textrm{STD}=\sqrt{\sigma^{2}+\frac{1}{\lambda^{2}}}
\label{eq:exgauss_std}
\end{equation}

The parameters $\mu$, $\sigma$ and $\lambda$ obtained are shown in table~\ref{tab:fit_parameters_exgauss}.

\begin{table}[h!]
  \begin{center}
    \begin{tabular}{c|c|c|c}
      Ch & $\mu$ & $\sigma$ & $\lambda$\\
      \hline
      HG & $480.2\pm5.6$ & $95.8\pm7.2$ & $(24.82\pm0.96)~10^{-4}$\\
      LG & $120.3\pm1.7$ & $25.3\pm2.2$ & $(98.6\pm4.6)~10^{-4}$\\
    \end{tabular}
	\caption{Parameters $\mu$, $\sigma$ and $\lambda$ from fit with equation~(\ref{eq:exgauss}) for HG and LG channels.}
	    \label{tab:fit_parameters_exgauss}
  \end{center}
\end{table}       

\begin{figure}[h]
  \centering
  \includegraphics[width=.7\textwidth]{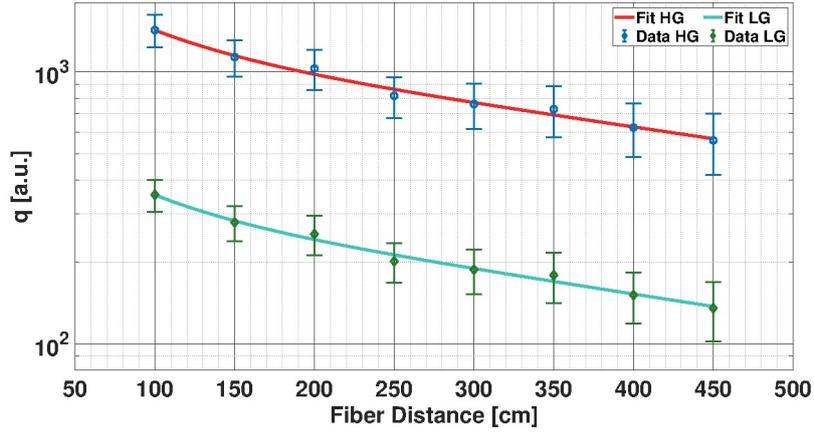}
  \caption{Attenuation of the muon charge signal along the scintillation bar for HG (blue dots) and LG (green dots). The red and green curves are fit function according to equation~(\ref{eq:double_ex}). The functional shape of the attenuation is similar as it depends mainly on the optical properties. The different absolute scale reflects the different channel gains.}
\label{fig:attenuation:curve}
\end{figure}

The values for the mean muon charge of table~\ref{tab:fit_parameters_exgauss} are averages from measurements at different positions. The analysis of the same data, but now for every position of the hodoscope separately, leads to the characteristic light attenuation curves. As it is explained in~\cite{Platino:2011,Sanchez:2010}, the double exponential function given in equation~(\ref{eq:double_ex}) is a correct model to fit the data points.
\begin{equation}
A_{tt}(x)=A_0 \, (a_f \, e^{- \frac{x}{\lambda_{1}}}+(1-a_f) \, e^{- \frac{x}{\lambda_{2}}})
\label{eq:double_ex}
\end{equation}
Here $A_{tt}$ is the average signal charge as a function of distance $x$ along the bar, $A_{0}$ is a scale factor related to electronic gain, $a_{f}$ is the relative fraction of two exponential functions with attenuation constants $\lambda_{1}$ and $\lambda_{2}$. The data of figure~\ref{fig:attenuation:curve} was fitted with equation~(\ref{eq:double_ex}) and the parameters obtained are shown in table~\ref{tab:fit_parameters_double_exp}
 
\begin{table}[h!]
  \begin{center}
    \begin{tabular}{c|c|c|c|c}
      Channel & $A_0$ & $a_f$ & $\lambda_1$ & $\lambda_2$ \\
      \hline
      HG & $3023\pm172$ & $0.55\pm0.02$ & $57.9\pm4.9$ & $509\pm16$\\
      LG & $786\pm48$ & $0.55\pm0.02$ & $53.8\pm4.4$ & $479\pm13$\\
    \end{tabular}
	\caption{Double exponential fit parameters.}
	    \label{tab:fit_parameters_double_exp}
  \end{center}
\end{table}

The parameters $a_{f}$, $\lambda_{1}$ and $\lambda_{2}$ depend on the light attenuation and the SiPM response, thus they reflect the properties of the optical fiber, the scintillation bars, the optical couplings and the SiPM pixels. As these factors are identical for HG and LG channels, one expects similar values for the fit parameters $a_{f}$, $\lambda_{1}$ and $\lambda_{2}$. The shape of the attenuation curve depends on these three parameters and both curves have a similar shape as can be seen in figure~\ref{fig:attenuation:curve}. However, the scale factor $A_{0}$ reflects the channel gain. Thus, the ratio of parameter $A_{0}$ for the two channels (HG and LG) is close to a factor 4 as set by design. This proves that HG and LG channels implement complementary behaviour and response, but at different designed operation range. 

\subsection{Linearity characterization}
\label{subsec:Linearity_characterization_setup}

\subsubsection{Methodology}
\label{subsubsec:Methodology}
For the determination of the muon density it is essential that the HG and LG outputs are linear within their full dynamic ranges. In system theory, linearity is defined by the principle of superposition, i.e.\ the system response to superpositioned input (\textit{x} + \textit{y}) is equal to the sum of the individual responses to input \textit{x} and \textit{y} as expressed in equation~(\ref{eq:linearity}):

\begin{equation}
f(A(x+y))=A \cdot f(x)+A \cdot  f(y)
\label{eq:linearity}
\end{equation}

Considering a constant frequency response along the entire spectrum, the task is to determine the gain $A$ and the dynamic range where this gain is constant. This requires measuring both, the input and output of the system. In our case, the ADC channels adds up the signals of 64 pixels of the SiPM. In the ideal case, each pixel \textit{i} is added with an unique gain \textit{G}. Thus, considering each pixel separately and following the superposition principle the output $y_{i}(t)$ is: 

\begin{equation}
y_{i}(t)=G \cdot x_{i}(t)
\label{eq:integrator_function}
\end{equation}

Where $y_{i}(t)$ is the output signal for pixel $i$, $x_{i}(t)$ is the input signal of pixel $i$ and $G$ is the gain. Assuming that the ADC channels are a linear system and considering equation~(\ref{eq:linearity}) with $A=1$, the output considering all signal input is:

\begin{equation}
y_{o}(x_{1}(t)+x_{2}(t)+x_{3}(t)+...+x_{64}(t))=G \cdot x_{1}(t)+G \cdot x_{2}(t)+G \cdot x_{3}(t)+...+G \cdot x_{64}(t)
\label{eq:integrator_linearity}
\end{equation}

Where $y_{o}$ is the total output signal of the ADC channels (sum of the 64 pixel signals), $x_{i}(t)$ are the input signals on each pixel and $G$ is the common gain for all pixels. On the other hand, the arithmetic sum of the individual signals at the output of the ADC channels (see equation~(\ref{eq:integrator_function})) is defined as $y_{s}(t)$.

\begin{equation}
y_{s}(t)=y_{1}(t)+y_{2}(t)+...+y_{64}(t)
\label{eq:integrator_arith_sum}
\end{equation}

As the ADC channels operation mode is charge measurement, the dynamic range characterization is made as a function of charge. Consequently, $Q_{i}$ is defined as the output charge measurement when there is a signal on channel $i$. $Q_{o}$ is defined as the output charge measurement when there are signals on all channels. Under these definitions, the equation~(\ref{eq:integrator_linearity}) is:

\begin{equation}
Q_{o}=Q_{1}+Q_{2}+...+Q_{64}
\label{eq:integrator_charge_relation}
\end{equation}

Equations (\ref{eq:integrator_arith_sum}) and (\ref{eq:integrator_charge_relation}) holds for perfect system linearity. The non-linearity NL (expressed in \%) is obtained by evaluating the relative error between the arithmetic sum of the individual charge and the analog sum:

\begin{equation}
NL=\frac{Q_{o}-(Q_{1}+Q_{2}+...+Q_{64})}{(Q_{1}+Q_{2}+...+Q_{64})} \cdot 100=\frac{Q_{o}-Q_{s}}{Q_{s}} \cdot 100
\label{eq:nl}
\end{equation}

Therefore, it is possible to obtain the total linearity curve simply by measuring the output charge for various levels of $Q_{i}$.
 
The setup described in the next section allows to illuminate each SiPM pixel individually by an array of 64 LEDs. For each charge value $Q_{i}$, different numbers of pixels can be excited to reach different values of $Q_{o}$ (from 1 to 64 pixels). This is called a test run and implies two steps. In a first step, each pixel $i$ is illuminated individually and the charge $Q_{i}$ at the output is recorded for several shots. In a second step, the pixels of the previous step are illuminate simultaneously and the output charge $Q_{o}$ is determined. The non-linearity (NL) is then obtained with equation~(\ref{eq:nl}) as:

\begin{equation}
NL=\frac{Q_{o}-\Sigma Q_{i}}{\Sigma Q_{i}} \cdot 100
\label{eq:nl_2}
\end{equation}

The NL values are determined for different values of $Q_{i}$ with different numbers of excited pixels.

\subsubsection{Setup}
\label{subsubsec:Linearity_setup}
\begin{figure}[h!]
	\centering
	\includegraphics[width=0.9\textwidth]{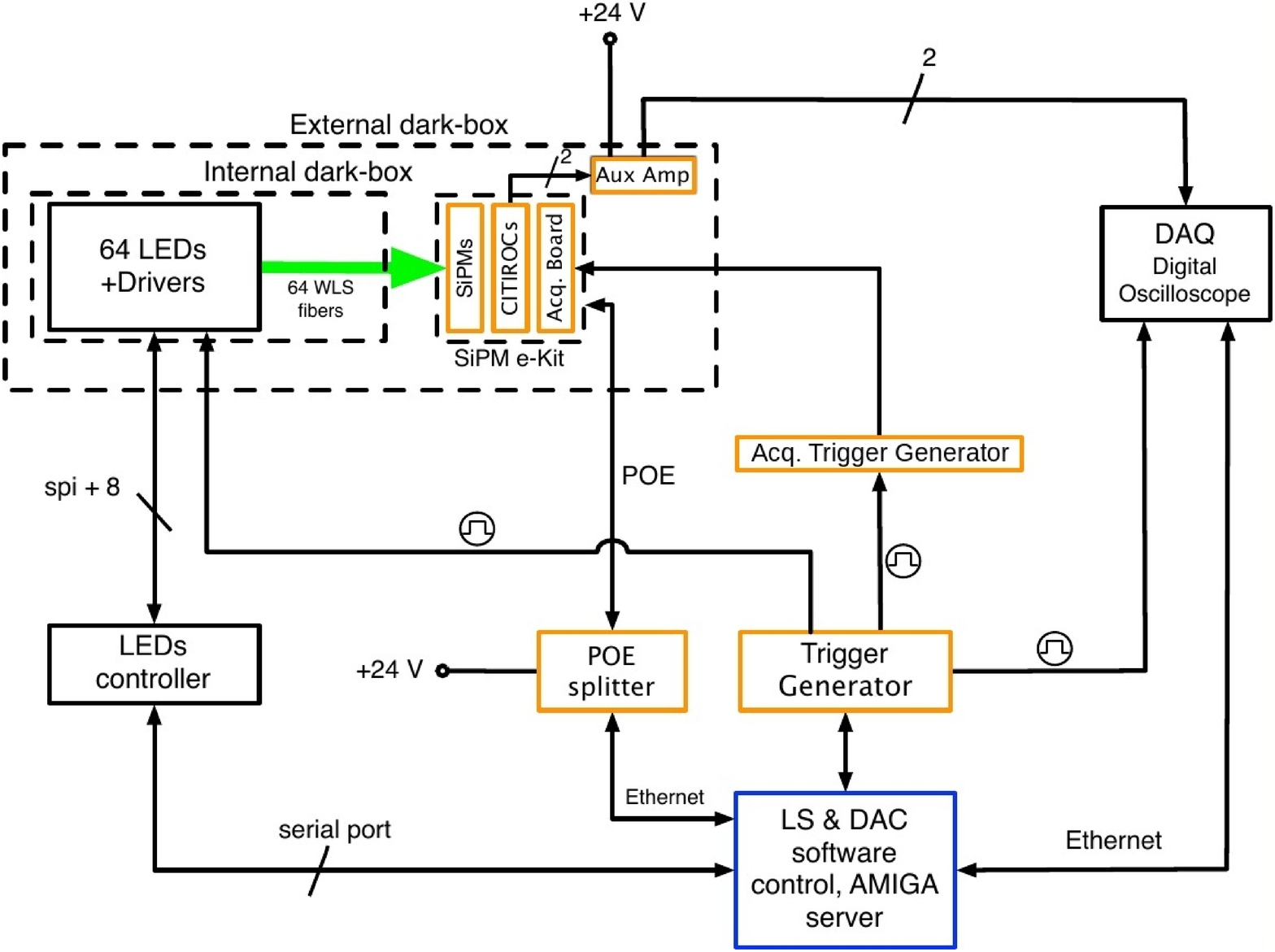}
	\caption{Setup for characterization of linearity. The outermost dashed line delimits the components inside the dark box: the 64 channel LED array (left) distributes the light via 64 fibers to the SiPM (right). Additional components for triggering are indicated by orange boxes. The computer for DAQ and test control (blue box) executes original AMIGA software routines.}
	\label{fig:darkbox_setup_scheme}
\end{figure}

For the linearity characterization, a 64 channel light source illuminates the SiPM pixels and generates individual input signals at the ADC channel inputs (see figure~\ref{fig:darkbox_setup_scheme}). 

This light source has been developed primarily for the MPPMT characterization and system test of the previous AMIGA electronics~\cite{Suarez:2011}.
Each LED has its own analog driver; the amplitude of each driver is defined by an individual 12-bit DAC steered from a common microcontroller which can be programmed through a serial interface from the control computer. The value set for this DAC is called LSDAC (value). Each LED is connected by the same type of WLS optical fiber as used in the AMIGA modules to the corresponding SiPM pixel. The flexible design allows illumination of SiPM pixel simultaneously or sequentially with a wide range of amplitudes. Design details and explanation of the internal structure are given in~\cite{Suarez:2011}. The setup is complemented by the AMIGA SiPM electronics (e-kit) inside the light-tide darkbox and auxiliar electronics for synchronous distributing a trigger signal to the light source and the electronics.

Due to the different light couplings and the varying LED characteristics, individual channels provide different pulse charges for the same LSDAC values. It is thus necessary to determine in a pre-test the individual LSDAC values per pixel needed to obtain the same charge. It is important to note that these values are not used as a reference, but they are used as initial values to optimize the linearity measurement. 
To obtain these values, the AMIGA SiPM electronics under test was put inside the darkbox, coupled to the LEDs, and calibrated using the method explained in~\cite{Hampel:2017}. In this way, the electronics is configured as if it was installed at the Pierre Auger Observatory site.     

As the binary channels (CITIROCs) and ADC channels measure in parallel, one analog output per CITIROC is connected to an independent auxiliary analog signal conditioning board (one per CITIROC) and from there, to the DSO\footnote{Digital Storage Oscilloscope.} WavePro 715Zi~\cite{DS-LECROY-WAVEPRO} from LeCroy Corporation to acquire the SiPM pulse of each pixel. With this setup, individual pulses per SiPM pixel have been acquired. A light source sweep was performed on all 64 SiPM pixels for LSDAC values within the range of 1200 to 2450, in steps of 5. 1000 events per LSDAC value were acquired and for each acquisition the pulse charge was determined.

\begin{figure}[h!]
	\centering
  	\includegraphics[width=0.7\linewidth]{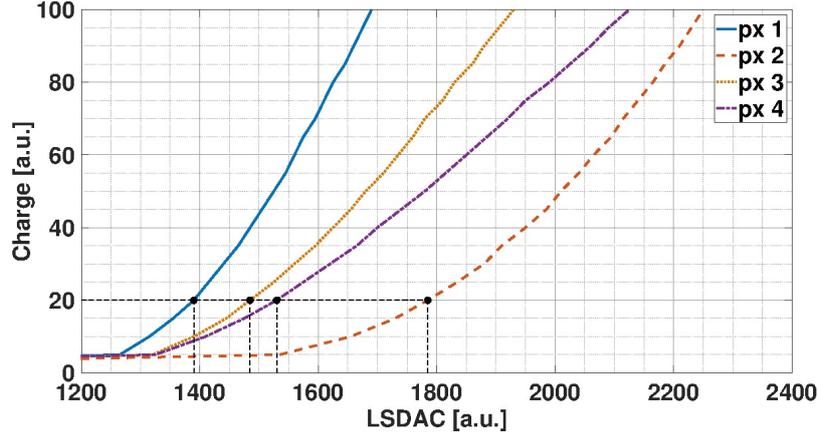}
  	\caption{Average charges of the pulses as function of LSDAC values for four different pixels. The dispersion between pixels can be observed. The black dots represent the individual LSDAC values needed for a charge of 20\,a.u.}
  	\label{fig:area_vs_dac}
\end{figure} 

The charge-vs-LSDAC values graph for four example pixels is shown in figure~\ref{fig:area_vs_dac}. As an example, the LSDAC values are indicated that yield a light level equivalent to a charge of 20\,a.u. With this different LSDAC values per pixel, a table for every pixels was composed with 20 LSDAC steps (64 values per step) that generate a charge varying from 5 to 100, in 5\,a.u.\ steps. These obtained LSDAC values are the ones used as initial values to perform the linearity curve measurement by using equation~(\ref{eq:nl}).

These LSDAC values are characteristic for every SiPM electronics, that means that the same process needs to be repeated for every electronics.

\subsubsection{Linearity measurement}
\label{subsubsec:Linealirity_measurement}
\begin{figure}[h]
\begin{subfigure}{.5\textwidth}
  \centering
  \includegraphics[width=1\linewidth]{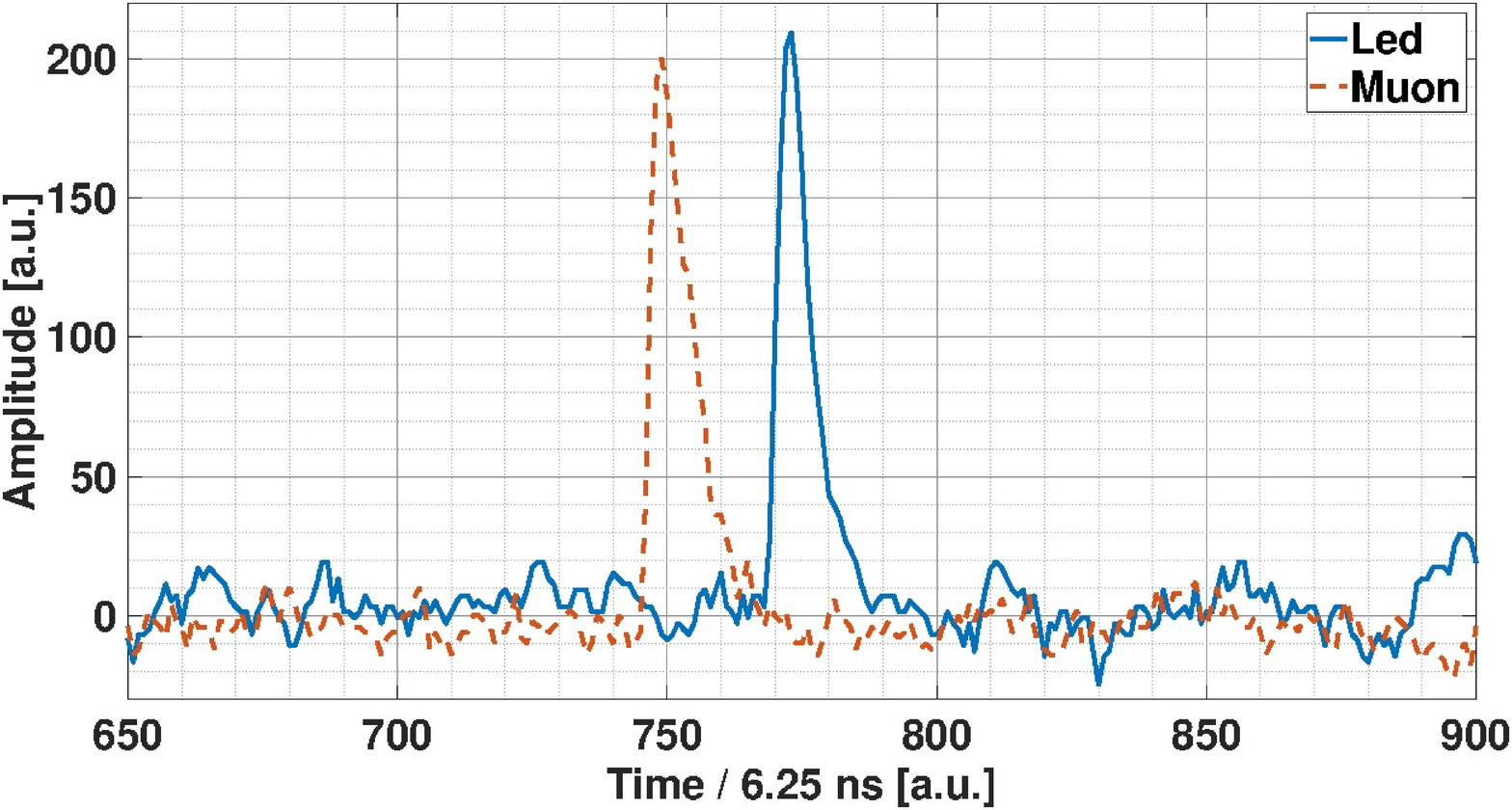}
  \caption{HG channel.}
  \label{sfig:hg_pulse_muon_and_pulse_led}
\end{subfigure}
\begin{subfigure}{.5\textwidth}
  \centering
  \includegraphics[width=1\linewidth]{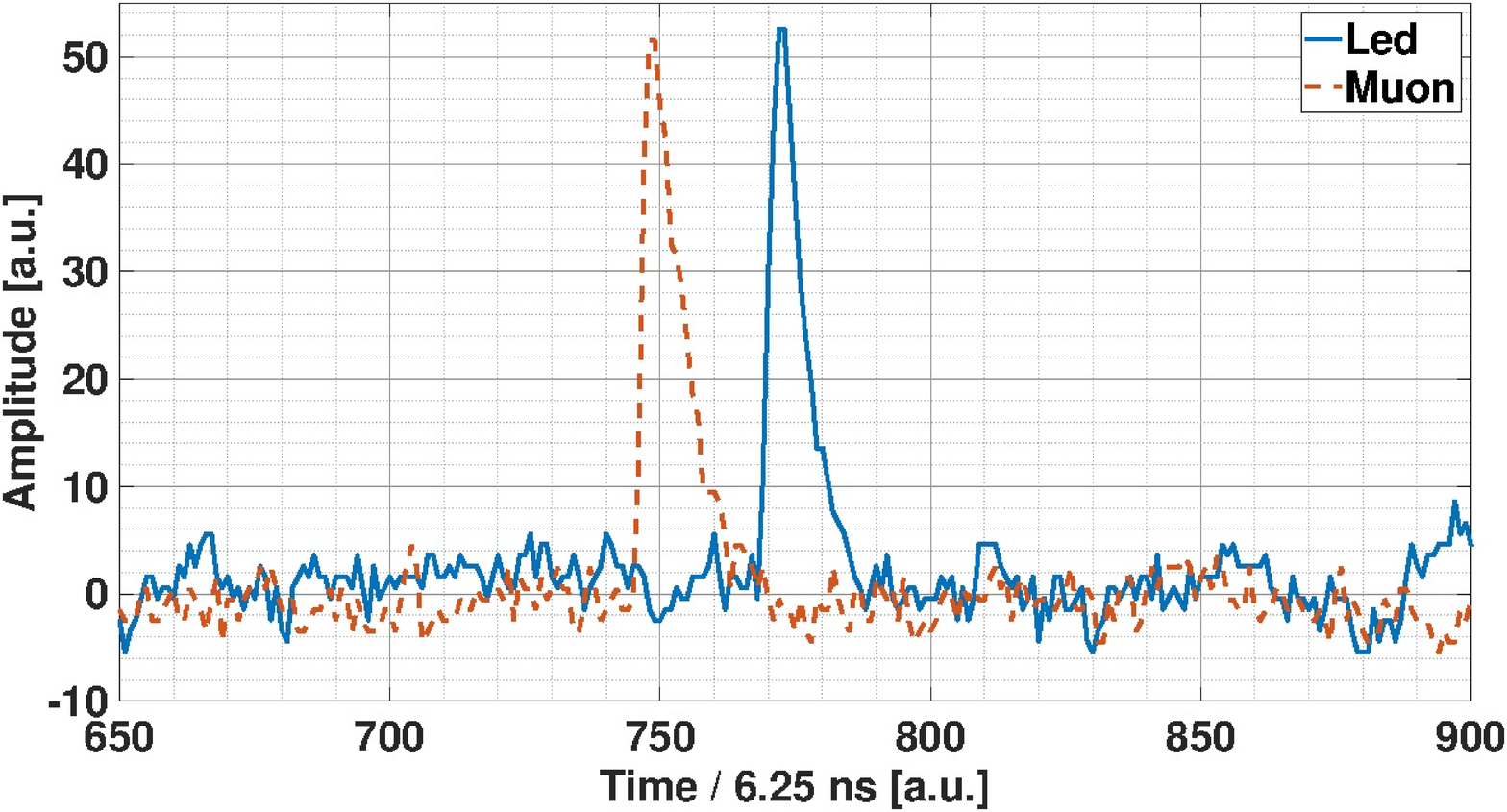}
  \caption{LG channel.}
  \label{sfig:lg_pulse_muon_and_pulse_led}
\end{subfigure}
\caption{Comparison of a pulse induced by a muon and a pulse initiated after finding the initial LSDAC values for the light source. The pulse shape is very similar for HG and LG.}
\label{fig:pulse_muon_and_pulse_led}
\end{figure}

Once the LSDAC initial values have been determined, the ADC channels dynamic range characterization is performed following the method explained in section~\ref{subsubsec:Methodology}. It consists of obtaining the linearity curve for both HG and LG channels, and finding the section of the curve that gives information about the dynamic range of particle measurement. The AMIGA acquisition system (server-client programs) has been used for the acquisition of the test events. No auxiliary board or DSO was required for this measurement. Figure~\ref{fig:pulse_muon_and_pulse_led} shows a pulse produced by the light source (in blue) compared to a pulse produced by a muon (in orange) for HG (a) and LG (b) channels. Both pulses have a very similar structure and shape.

Different programs are responsible for configuring the light source, for controlling the signal generator that triggers the light source, the electronics and the synchronization of data acquisition, and for the control of event readout and storage. This setup emulates the worst saturation condition, i.e.\ a situation where all light signals arrive at the photo-sensor at the same time bin. Therefore, the linearity limits obtained with this measurement are the worst case scenario of the whole system, being improved under real conditions where the light arriving to the photo-sensor is spread over several time bins. The procedure used to measure the curves is as follows:

\begin{enumerate}
\item Configure the light intensity for each channel (LSDAC values) individually and trigger one at a time. Record N events per channels and calculate the distribution of the pulse charge.
\item The mean value $Q_{i}$ of the distribution corresponds to the individual mean charge at the output when only the channel $i$ is excited. This is repeated for the desired number of inputs L (for example, eight inputs, two inputs per adder). 
\item With the same light intensity (i.e.\ the same LSDAC) as before, the L inputs are shot simultaneously and the same amount of N events are acquired. 
\item A charge histogram is made and the mean value of the total charge $Q_{o}$ is obtained. This value corresponds to the mean charge at the output when all L inputs receive a light signal.
\item Steps 1 to 4 are repeated for 20 different light intensities or LSDAC values.
\end{enumerate}

Despite the high number of events, the measurement lasts only about 5 minutes. During this time, the environmental conditions don't change significantly, thus e.g.\ temperature effects are negligible.

To cover the whole range of the HG and LG channels, multiple linearity curves have been obtained: for low charges at a reduced number of inputs (L=4, single adder) up to very high charges (L=64 or 16 per adder), where the saturation is seen as deviation from linearity. 

The linearity curve is obtained by measuring $Q_{o}$ as a function of $Q_{s}$, where $Q_{s}= \sum_{i=1}^{L} Q_{i}$. When the relative error between $Q_{o}$ and $Q_{s}$ is greater than 5\,\%, the ADC channels are considered to be saturated. This non-linearity can equivalently be defined as a deviation of 5\,\% or more from the identity function $y=x$.

The measurements obtained in terms of charge are converted to number of equivalent simultaneous muons by using the mean muon charge values from figure~\ref{fig:muon_charge_setup_1} of section~\ref{muon_events_charge} ($q_{avg}$).

\begin{figure}[h]
	\centering
	\includegraphics[width=.7\textwidth]{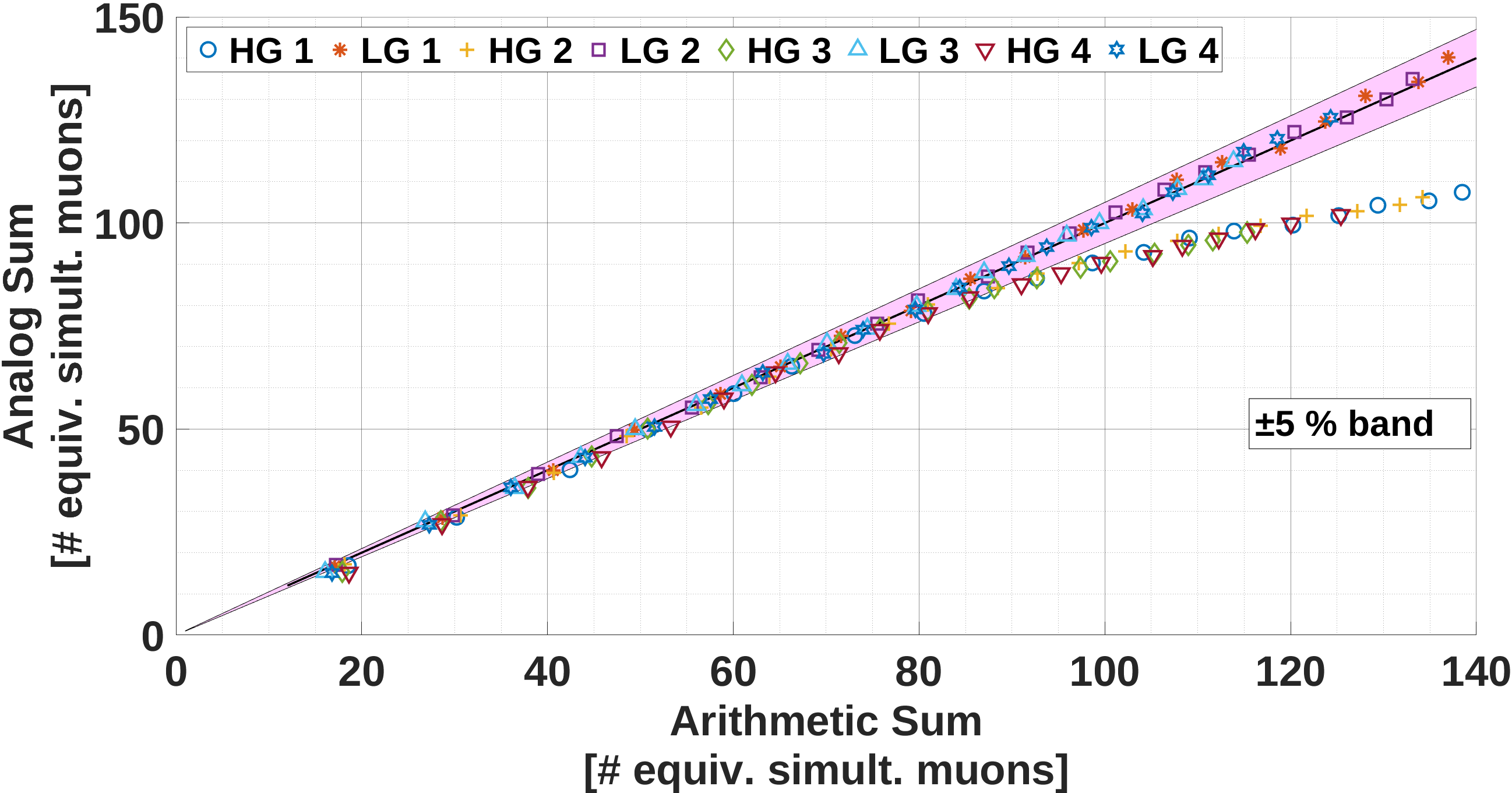}
	\caption{Linearity curves of the four adders. HG $n$ and LG $n$ are the linearity curve for adder $n$.}
	\label{fig:linearity_curve_per_adder}
\end{figure}

The first tests measured the linearity curve of each adder of the HG and LG setup (see section~\ref{subsec:Scheme overview}). Using the respective input channels of the adders (16 for each adder), the procedure described above was applied. The resulting plots of each adder for the HG and LG channels are shown in figure~\ref{fig:linearity_curve_per_adder}. Additionally, the identity function $y=x$ is shown as a reference. The saturation observed in the HG channel is expected due to its lower dynamic range, the LG channel does not present saturation. Within the limits of this test, the adders do not saturate individually, having a wide dynamic range for particle detection.

With additional tests, the complete linearity curves were found for HG and LG. As mentioned previously, in order to cover the entire range of admitted charges, several runs were made using a different number of inputs. For each new run, one more input was added for each of the four adders. In consequence, four inputs were added at a time, up to a total number of L = 60 inputs. 

\begin{figure}[ht]
  %\centering
  \begin{subfigure}{.5\linewidth}
   \centering
  \includegraphics[width=\textwidth]{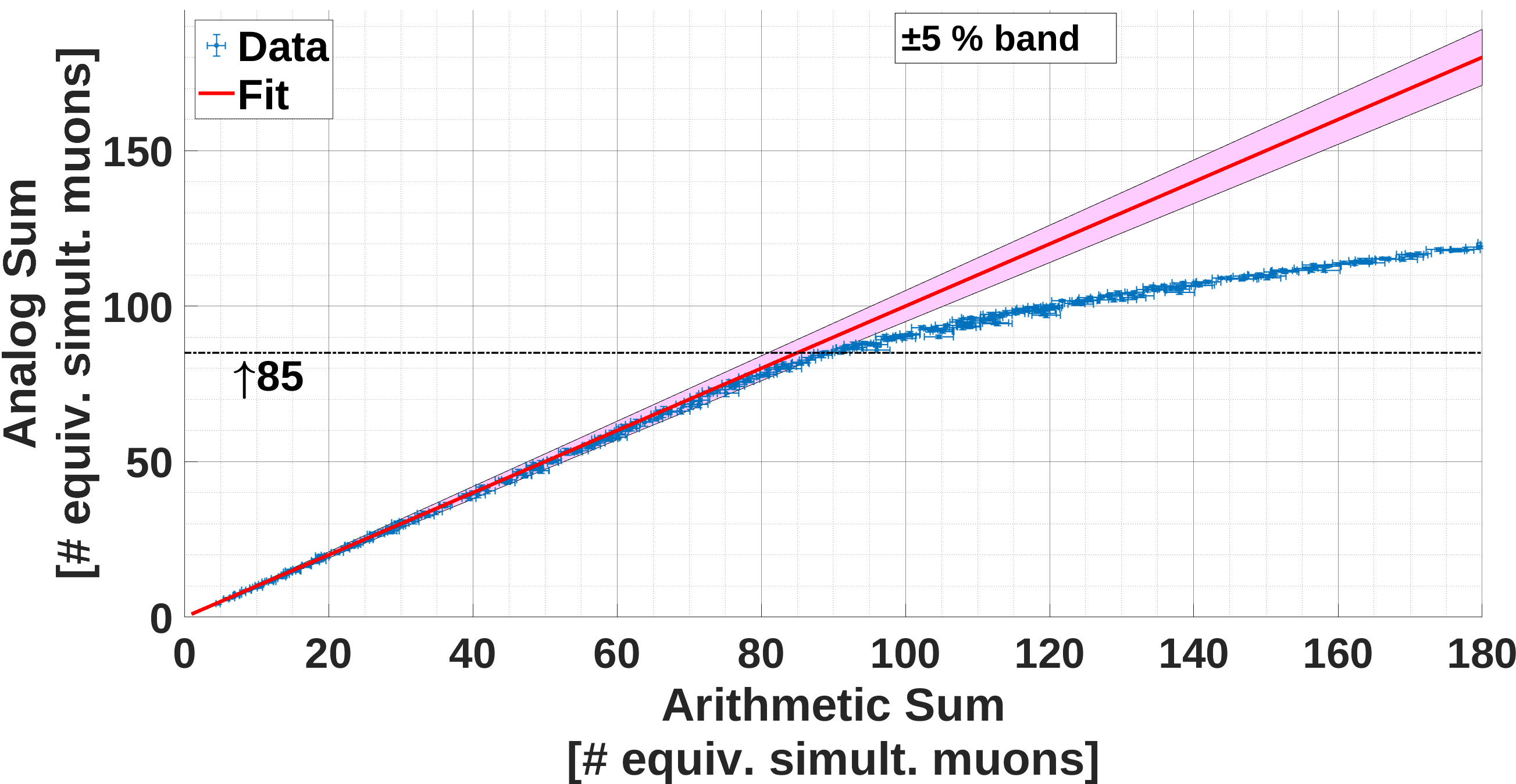}
  \caption{HG channel linearity curve.}
  \label{sfig:hg_linearity_curve}
  \end{subfigure}
  \begin{subfigure}{.5\linewidth}
  \centering
  \includegraphics[width=\textwidth]{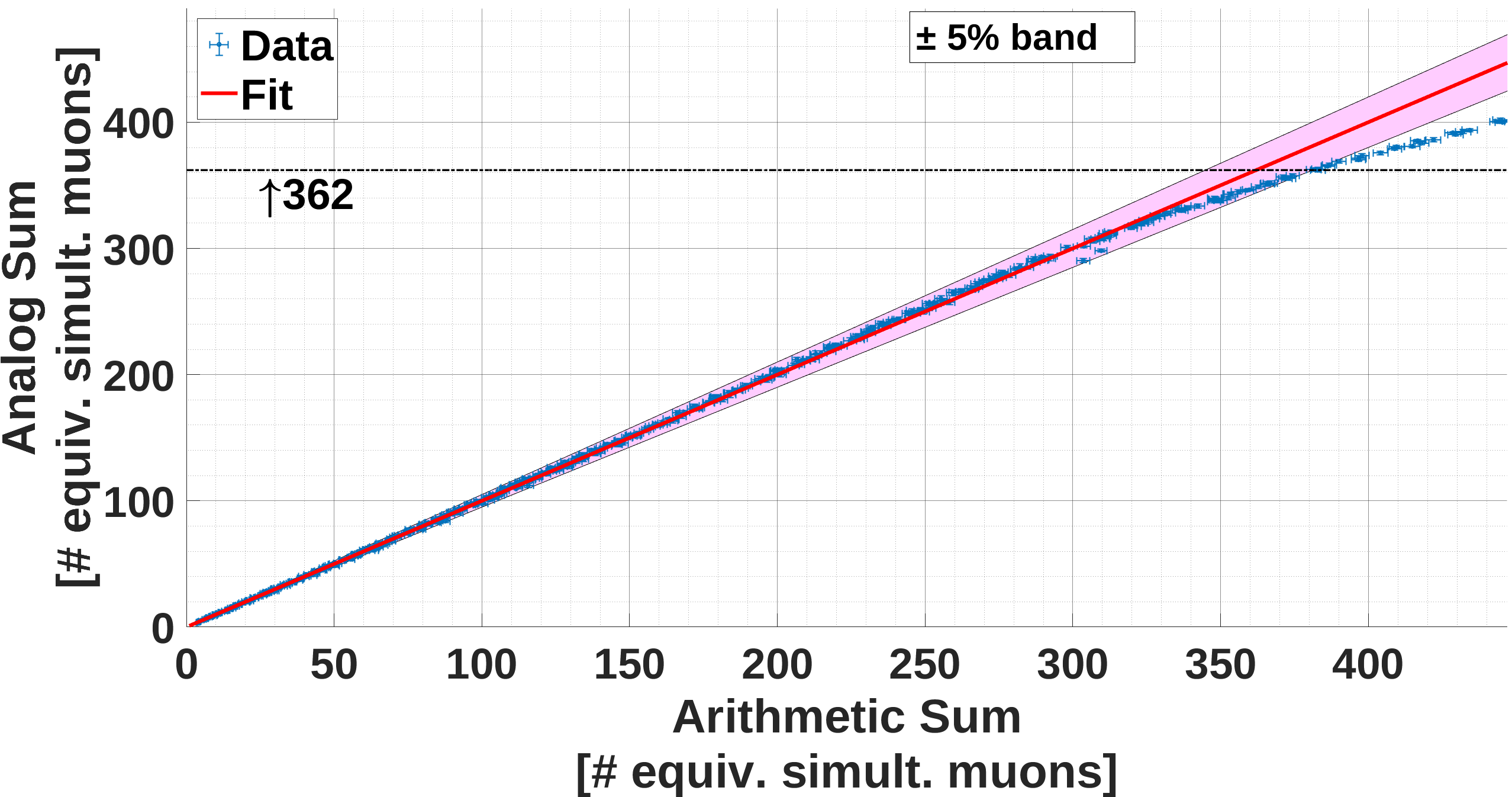}
  \caption{LG channel linearity curve.}
  \label{sfig:lg_linearity_curve}
  \end{subfigure}  
\caption{Linearity curves for HG (a) and LG (b) channels together with the identity function $y=x$ as a reference. The dashed line indicates the maximum muon value where the channel is linear within a 5\,\% deviation band.} 
\label{fig:linearity_curve}
\end{figure}

The complete linearity curves were obtained from the data of the sequence previously explained. Figure~\ref{sfig:hg_linearity_curve} shows the HG linearity curve. Saturation effects (deviation of $\geq$ 5\,\% from perfect linearity $y=x$) starts at a charge equivalent to 85 simultaneous muons. Above this equivalent charge value, the LG channels should be used. Figure~\ref{sfig:lg_linearity_curve} shows the LG linearity curve. Similarly, this channel starts to saturate from a charge equivalent to 362 simultaneous muons. It is important to note, that the minimum charge that the HG and LG channel can measure ($\approx$1\,muon) allows us
to cross-characterize the total charge measurement with the muons number obtained by the binary channels. 

The observed non-linearity is caused by saturation of the electronic amplifiers. The linearity of SiPMs is quite well known and might be explored in future works.

\subsection{On-site results}
\label{subsec:On-site results}
\begin{figure}[h]
	\centering
	\includegraphics[width=0.6\textwidth]{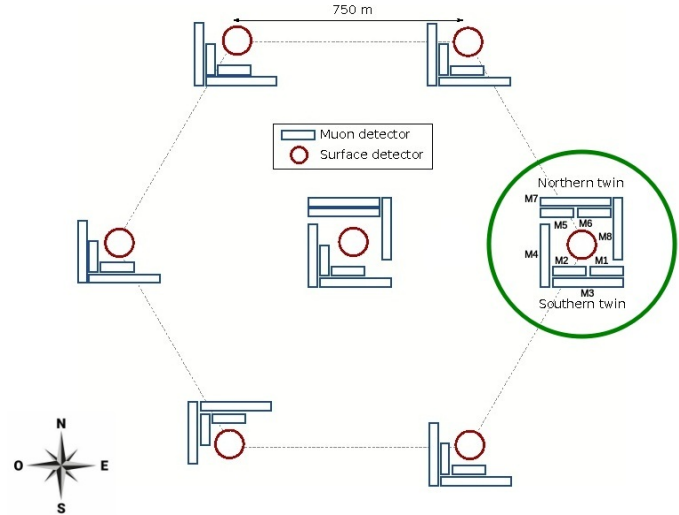}
	\caption{Sketch of the underground muon detector engineering array. Each circle represents a position of the SD detector and each rectangle represents a module of the UMD. The station with eight modules is used for this analysis (inside the green circle).}
	\label{fig:uc_amiga}
\end{figure}

Data from one UMD station of the engineering array (see figure~\ref{fig:uc_amiga}) have been analysed to verify its performance and validate the method of on-site characterization. 

The selected station is equipped with eight modules (twin-type arrangement) that have SiPM as photodetector and is fully operational since October 2016. As this position is the first station that was completely equipped with SiPM electronics, the largest amount of data was acquired and stored. Thus, this stations provides the highest statistic for performance studies of the new front-end. The analysis takes into account data from four modules of 5\,m$^2$ size (\# 1, 2, 5 and 6) and four modules of 10\,m$^2$ size (\# 3, 4, 7 and 8). The modules of the station are grouped in southern modules (\# 1 to 4) and northern modules (\# 5 to 8). The analysis considers only events where energy, arrival direction and impact point on ground of the EAS could be reconstructed from SD-750 data of surrounding stations of the infill.  

From those events, only showers with zenith angle $\theta \leq$ 45\textdegree and with energies $\geq$ 2 x 10$^{17}$\,eV were analysed~\cite{Wundheiler:2020,WundheilerICRC:2016}. The cut in zenith angle allows to minimize the attenuation effects and the statistical uncertainties due to the reduced scintillation-module detection area, while the cut in energy allows us to work in a regime of the full efficiency of the EAS array. No cut in distance to shower core is made. For the analysis of the binary channels data, corrections of the pile-up effect~\cite{Supanitsky2008} and the time synchronization with the SD data have been done. The pile-up effect occurs when a single segment or scintillator bar is impacted by two or more particles simultaneously, resulting in undercounting. Then, only events with muon numbers $\leq$ 85 and $\leq$ 362 measured by the binary channels were considered for the characterization of the HG and LG channels, respectively. Both cuts in muon numbers allow and ensure to analyse the ADC channels working within the linearity range (see section~\ref{subsubsec:Linealirity_measurement}). The number of events found in the different modules after applying these selection criteria are given in table~\ref{tab:events_char} for HG and LG.

\begin{table}[h!]
	\begin{center}
	\begin{tabular}{|c|c|c|}
	\hline
	\textbf{Events} & \textbf{HG} & \textbf{LG} \\ \hline
		M1 & 8043 & 8141 \\ \hline
		M2 & 7974 & 8077 \\ \hline
		M3 & 10013 & 10173 \\ \hline
		M4 & 10078 & 10240 \\ \hline
		M5 & 7970 & 8057 \\ \hline
		M6 & 7765 & 7853 \\ \hline
		M7 & 9249 & 9391 \\ \hline
		M8 & 10108 & 10291 \\ \hline
	\end{tabular}
	\caption{Number of events per module.}
	\label{tab:events_char}
	\end{center}
\end{table}

\begin{figure}[h!]
  %\centering
  \begin{subfigure}{.5\linewidth}
    \centering
  \includegraphics[width=\textwidth]{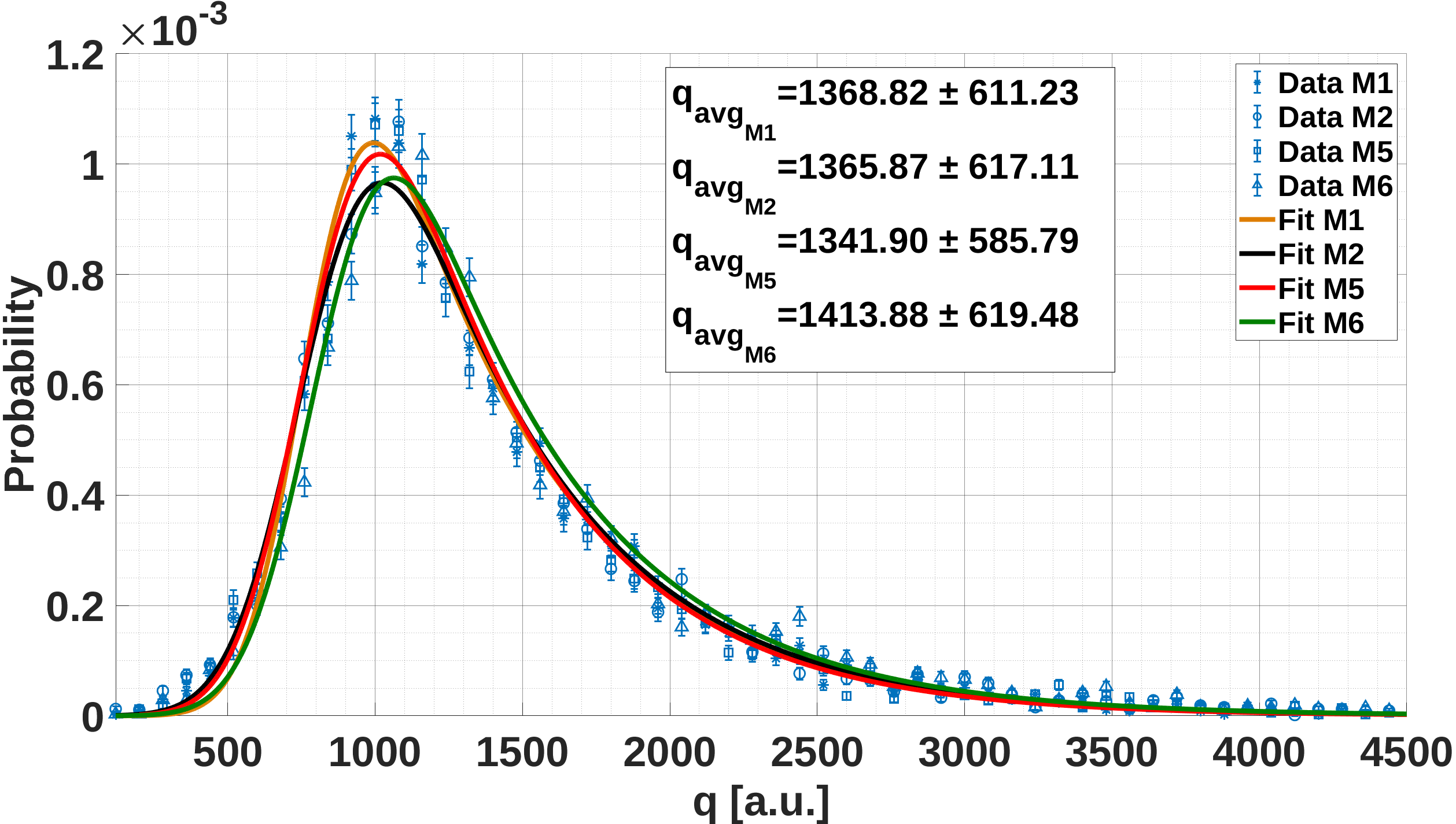}
  \caption{5\,m$^2$ modules.}
  \label{sfig:kt_hg_5m}
  \end{subfigure}
  \begin{subfigure}{.5\linewidth}
    \centering
  \includegraphics[width=\textwidth]{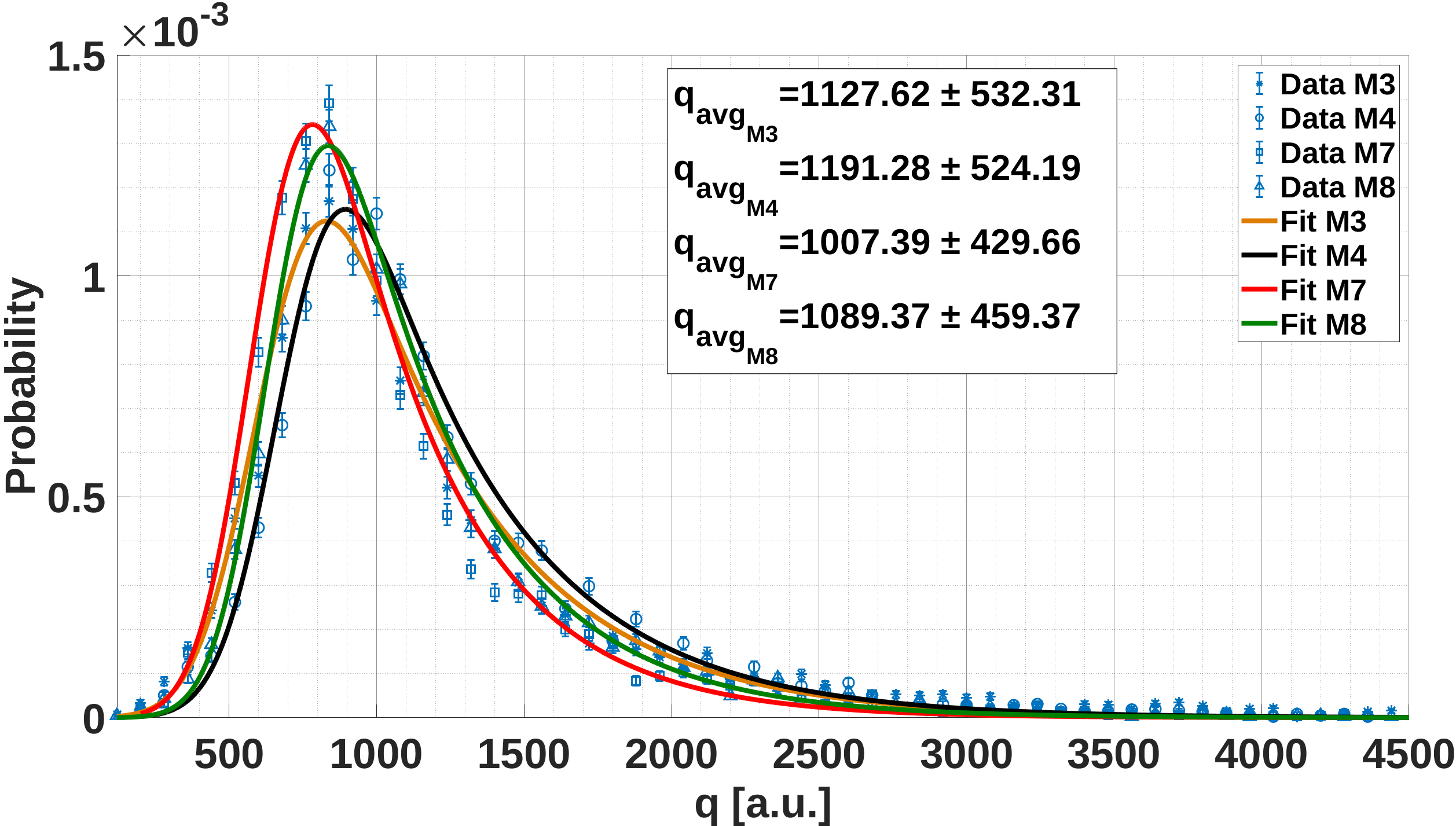}
  \caption{10\,m$^2$ modules.}
  \label{sfig:kt_hg_10m}
  \end{subfigure}  
\caption{HG channel muon charge histograms for different modules. The data set corresponds to the period from May 2017 to December 2018.}
\label{fig:kt_hg}
\end{figure}

\begin{figure}[h!]
  %\centering
  \begin{subfigure}{.5\linewidth}
  \centering
  	\includegraphics[width=\linewidth]{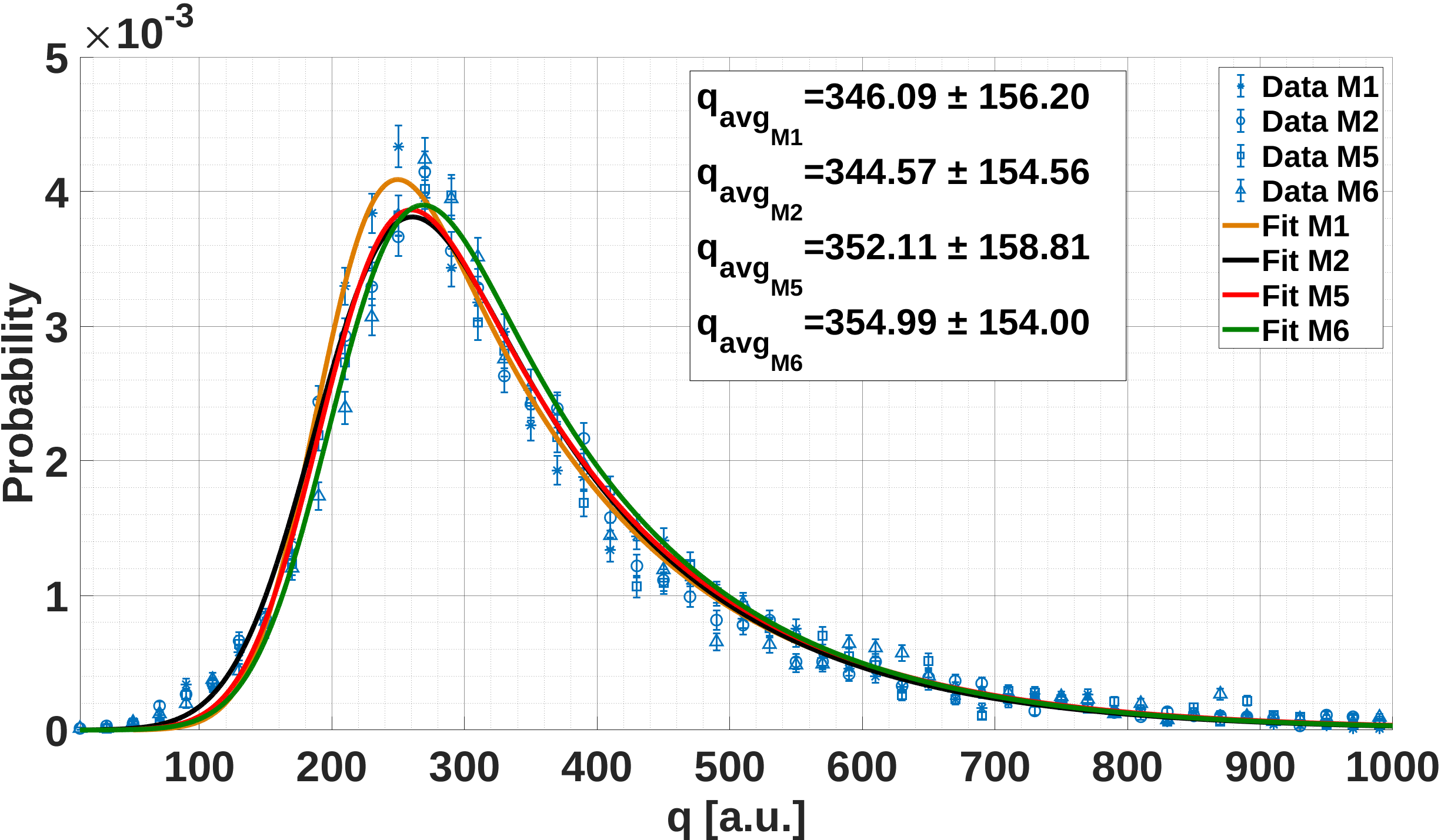}
  	\caption{5\,m$^2$ modules.}
  \label{sfig:kt_lg_5m}
  \end{subfigure}
  \begin{subfigure}{.5\linewidth}
  \centering
  	\includegraphics[width=\linewidth]{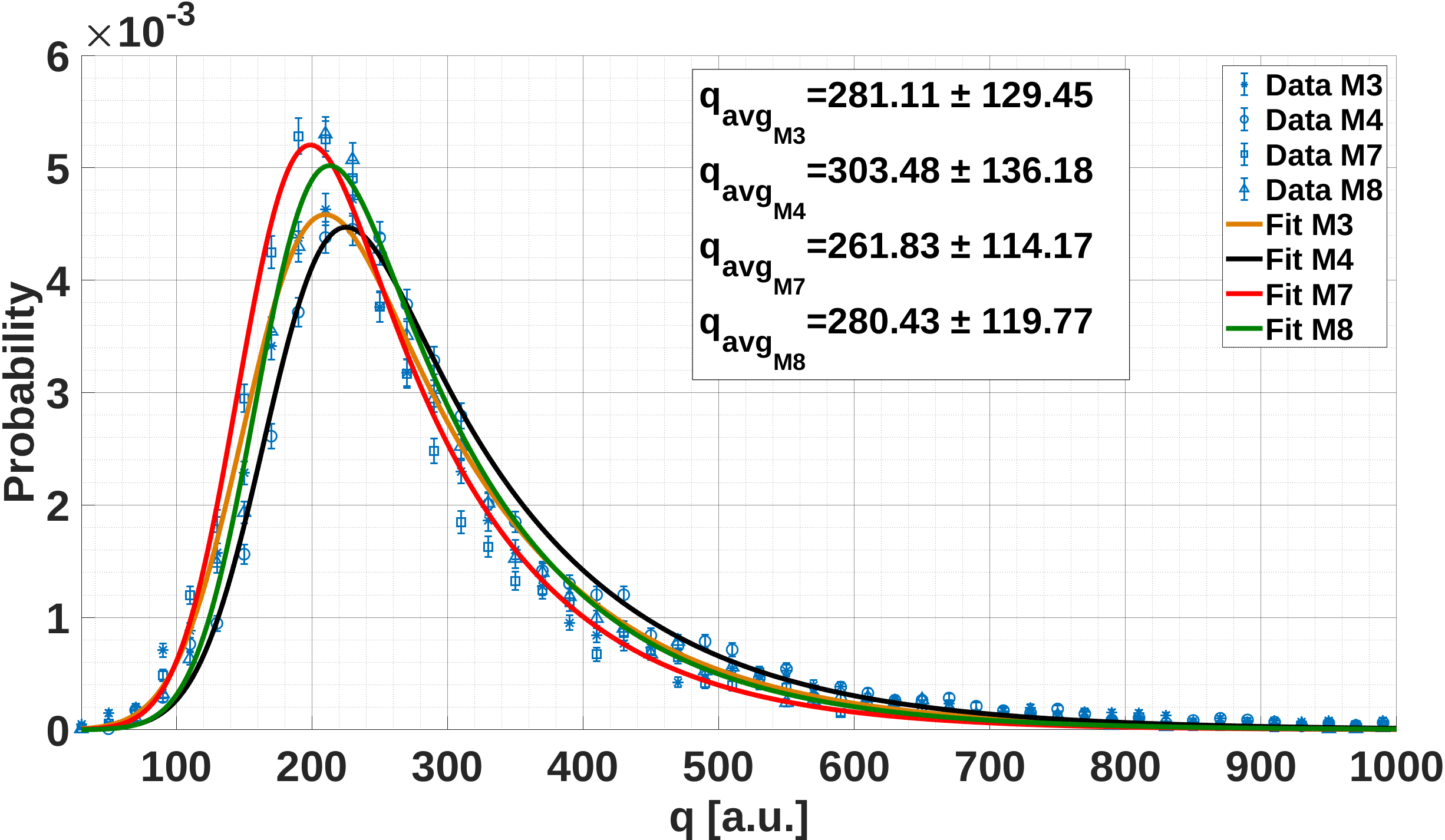}
  	\caption{10\,m$^2$ modules.}
  \label{sfig:kt_lg_10m}
  \end{subfigure}  
\caption{LG channel muon charge histograms for different modules. The data set corresponds to the period from May 2017 to December 2018.}
\label{fig:kt_lg}
\end{figure}

The charge histograms of one muon for the HG and LG channels of the eight modules were determined by making the quotient between the measured charge and the number of muons obtained by the binary channels in the same event. Each histogram was fitted with equation~(\ref{eq:exgauss}) of the exponentially modified gaussian distribution as explained in section~\ref{muon_events_charge}. The results are shown in figure~\ref{fig:kt_hg} for the HG channel and in figure~\ref{fig:kt_lg} for the LG channel, respectively (separate for the 5\,m$^2$ modules (a) and the 10\,m$^2$ modules (b)). 

From the figures, it can be seen that the mean muon charge value of the 5\,m$^2$ modules is higher than the one of the 10\,m$^2$ modules (for both HG and LG). This can be explained with the higher light attenuation in the larger modules. On average, the modules behave as if the muon impinges on the middle of the scintillation bar. Light needs to travel in 5\,m$^2$ modules on average 1\,m to the fiber end (2\,m long scintillator bars), but 2\,m in 10\,m$^2$ modules (4\,m long bars). Thus, the longer light path in larger modules lead to higher light attenuation. In consequence, the average charge measured in larger modules is lower.

\begin{figure}[h!]
\begin{subfigure}{.5\textwidth}
  %\centering
  \includegraphics[width=0.98\linewidth]{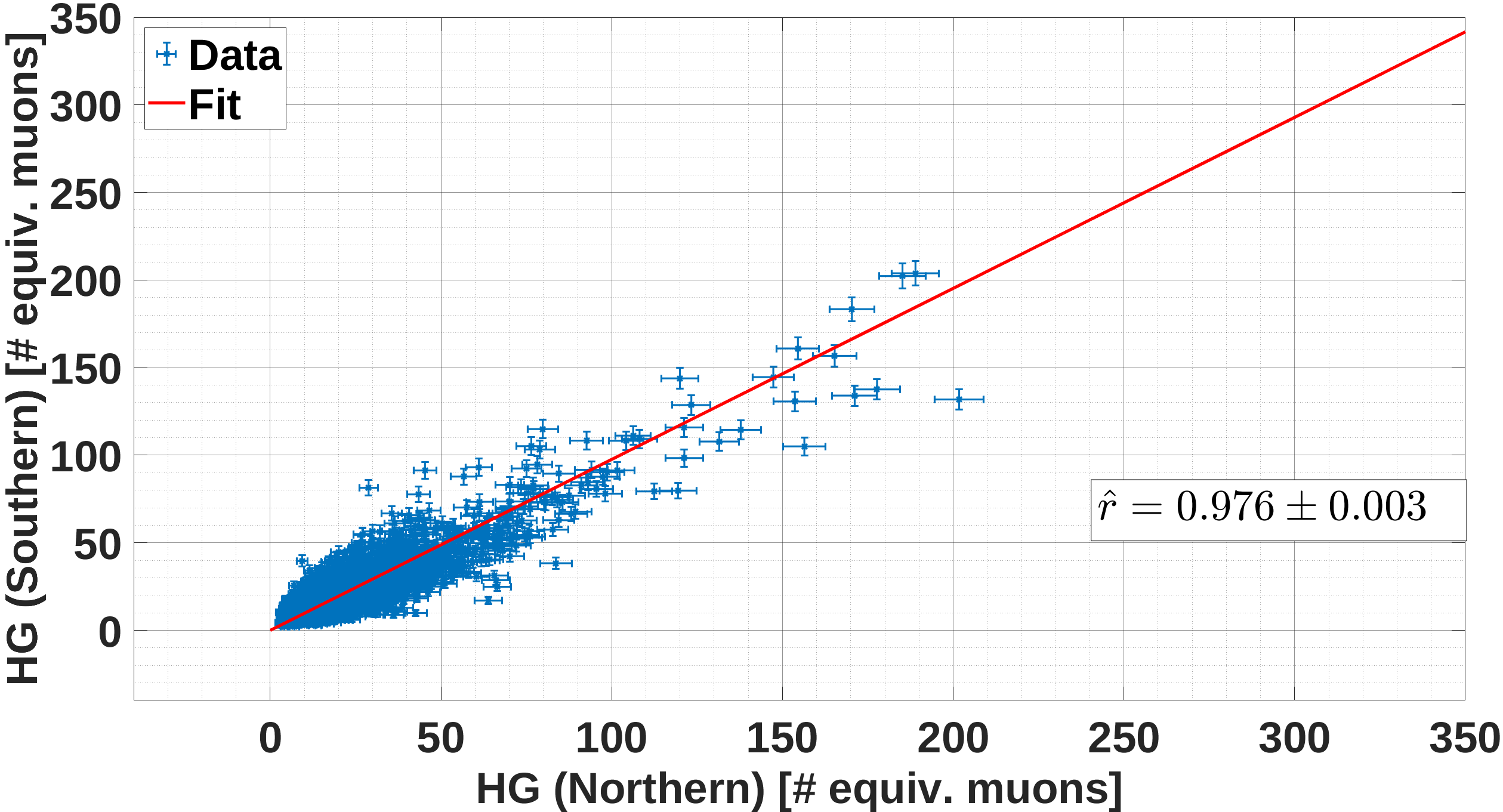}
  \caption{HG channels.}
  \label{sfig:hg_kt_comp}
\end{subfigure}
\begin{subfigure}{.5\textwidth}
  \centering
  \includegraphics[width=0.98\linewidth]{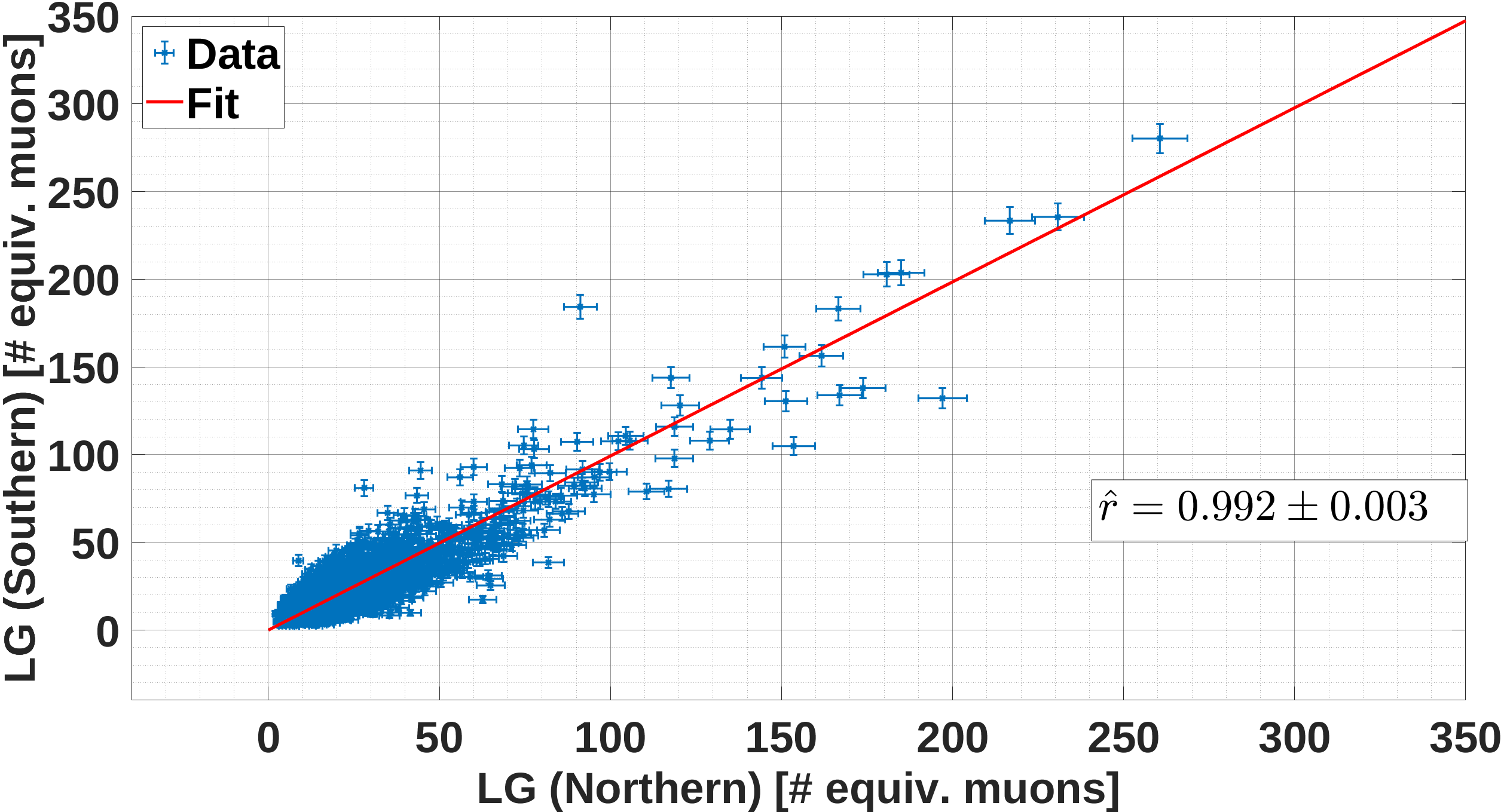}
  \caption{LG channels.}
  \label{sfig:lg_kt_comp}
\end{subfigure}
\caption{Comparison of the muons measured in the southern modules versus northern modules for the HG (a) and LG (b) channels. The red line represents a strait line fit of the ratio.}
\label{fig:kt_comp}
\end{figure}

The method of muon determination by total charge in HG and LG channels was verified by comparing groups of detector modules in the southern and in the northern part. Figure~\ref{fig:kt_comp} compares the number of muons recorded in the southern part with those measured by northern part modules. The ratio of number of muons measured, $\hat{r}$, can be estimated by equation~(\ref{eq:relacion_correlacion})~\cite{WundheilerICRC:2016,Wundheiler:2020}. 

\begin{equation}
\hat{r}=\frac{\sum^{N}_{i=1}(\mu_{M1_{i}}+\mu_{M2_{i}}+\mu_{M3_{i}}+\mu_{M4_{i}})}{\sum^{N}_{j=1}(\mu_{M5_{j}}+\mu_{M6_{j}}+\mu_{M7_{j}}+\mu_{M8_{j}})}
\label{eq:relacion_correlacion}
\end{equation}
Here $\mu_{M_{i}}$ is the muon equivalent value measured by the module $M$ in the event $i$ and $N$ is the total number of events in the data set. $\hat{r}$ is a maximum likelihood estimator.

We have excluded events with a shower core closer than 200\,m away from the analysis as in those events the high muon density gradient influences the muon counting in the 20\,m separation between the groups of modules significantly. Such a cut in distance also limits the fluctuations of the muon density. However, no cuts in zenith angle and energy were applied and a total of 9809 and 9818 events were found in the HG and LG data set, respectively.

The ratios line fits show slopes of $0.976\pm0.003$ (HG) and $0.992\pm0.003$ (LG). A ratio close to unity proves an equal behaviour of both groups of modules and the consistency of the ADC channels design.

\begin{figure}[h!]
\begin{subfigure}{.5\textwidth}
  %\centering
  \includegraphics[width=.98\linewidth]{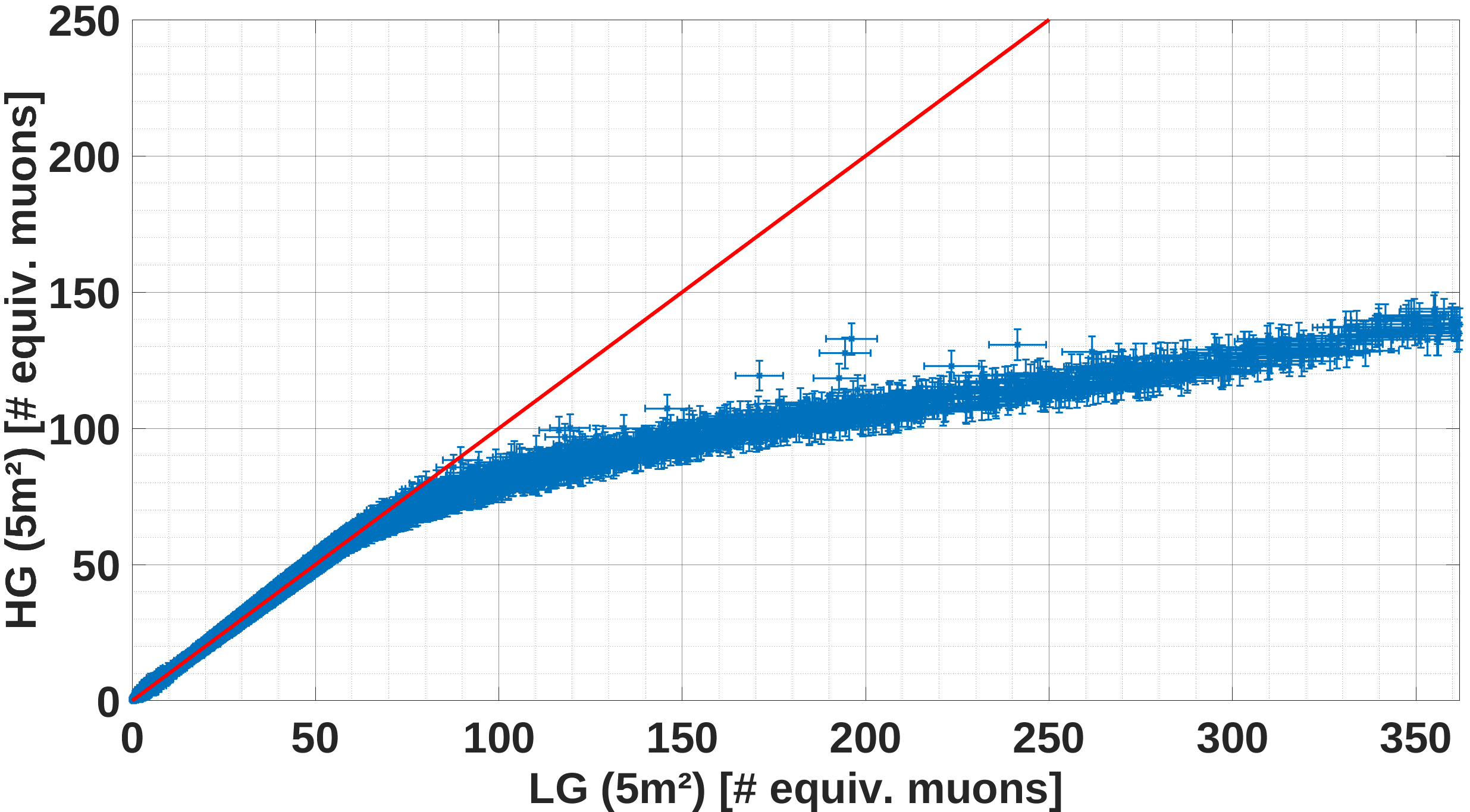}
  \caption{5\,m$^{2}$.}
  \label{sfig:kt_lg_vs_hg_5m}
\end{subfigure}
\begin{subfigure}{.5\textwidth}
  \centering
  \includegraphics[width=.98\linewidth]{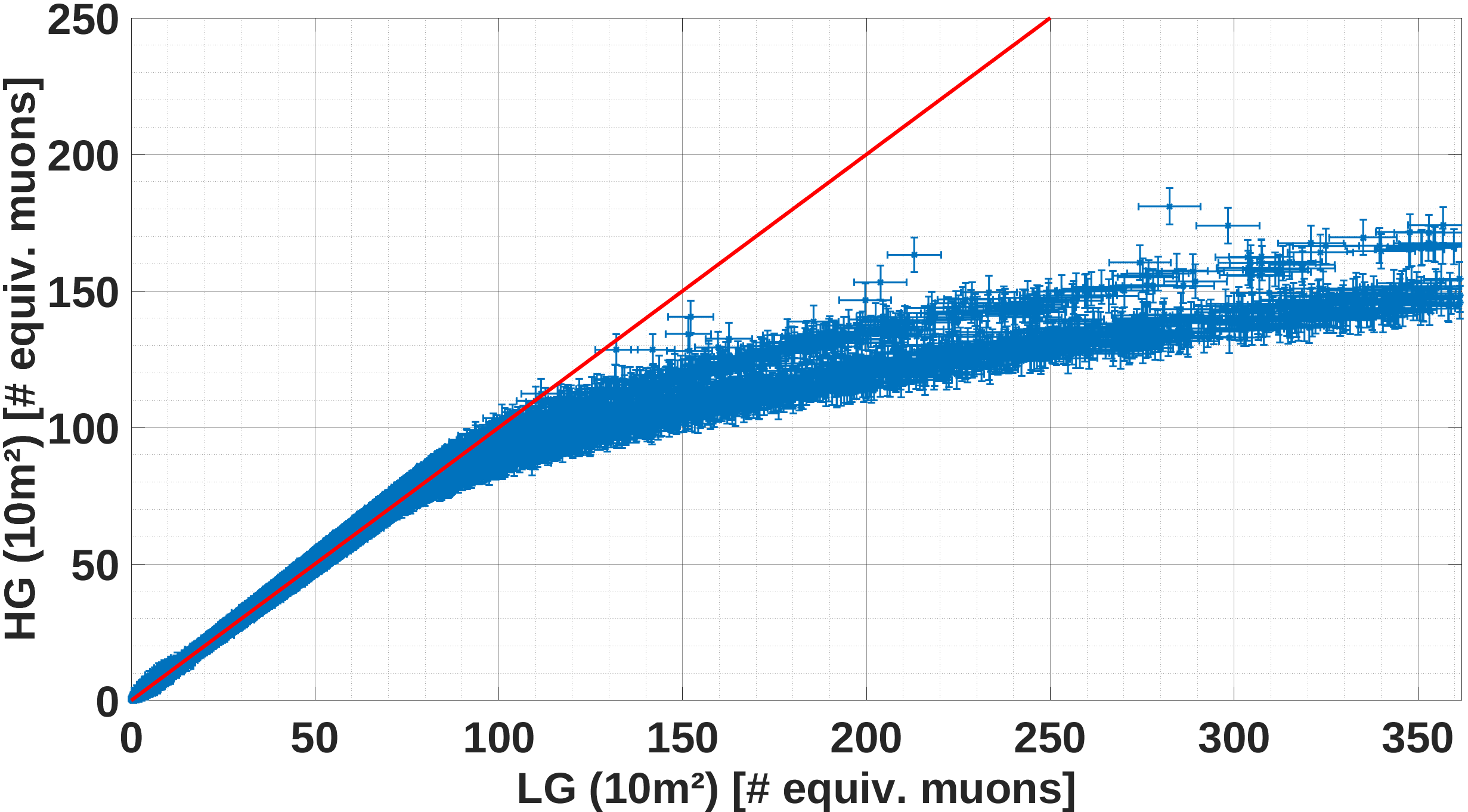}
  \caption{10\,m$^{2}$.}
  \label{sfig:kt_lg_vs_hg_10m}
\end{subfigure}
\caption{Comparison plot of HG vs LG for 5\,m$^2$ and 10\,m$^2$ modules. The red line represents equal response of HG and LG ($y(x) = x$). The HG channel saturates earlier for 5\,m$^{2}$ modules than for the larger 10\,m$^{2}$ modules.}
\label{fig:kt_lg_vs_hg}
\end{figure}

Finally, figure~\ref{fig:kt_lg_vs_hg} compares the number of muons measured by the HG and LG channels for 5\,m$^2$ modules (a) and for 10\,m$^2$ modules (b). The data sets of these plots are not restricted by any cut. Thus, the analysis was applied to about 200 000 events. The operating ranges of the ADC channels can be observed and have upper limits higher than those obtained in section~\ref{subsubsec:Linealirity_measurement}. This behaviour is expected because the linearity curves were obtained under worst case condition (all light signals reach the SiPM at the same time bin). The analysis shows a saturation of the HG channel around 75 equivalent muons for the 5\,m$^2$ modules, but of around 90 - 100 equivalent muons for the 10\,m$^2$ modules. This difference can easily be explained by the higher light attenuation in the larger module, which shifts the saturation point.

\subsection{Power consumption measurement}
\label{subsec:power_consumption_measurement}
\begin{figure}[h!]
	\centering
	\includegraphics[width=.7\textwidth]{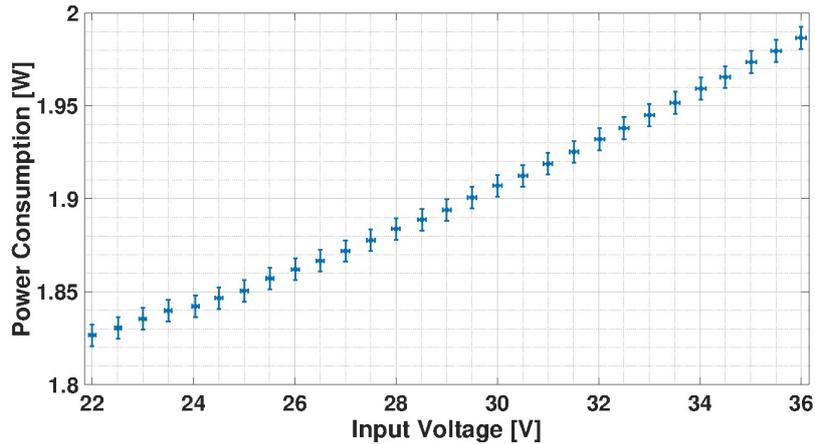}
	\caption{Power consumption versus front-end input voltage. All values are below 2\,W along all voltage values.}
	\label{fig:power:consumption_measurement}
\end{figure}

As mentioned in section~\ref{subsec:Power system}, a low power consumption is one of the main design goals. The power consumption was measured for an input voltage range from 22\,V to 36\,V (AMIGA battery voltage range: 23\,V to 32\,V~\cite{Cancio:2018}) with the SiPM board connected to the front-end board. In the measurement, we recorded simultaneously the input voltage with the multimeter U1231A~\cite{DS-U1231A} (60\,V full scale) and the input current with the multimeter U1242A~\cite{DS-U1242A} (100\,mA full scale). Figure~\ref{fig:power:consumption_measurement} shows the power consumption versus the input voltage. The power consumption increases with raising supply voltage from 1.83\,W to 1.99\,W, but it stays within the specification of maximal 2\,W for the front-end board combined with the SiPM. The slightly higher power consumption for higher supply voltage can be explained by the lower efficiency of the DC/DC regulator at higher input voltage.

\section{Conclusions}
\label{sec:Conclusions}
This paper describes the design and the implementation of the front-end electronics for AMIGA in its production phase. The new front-end shows a higher dynamic range (detection of above 362 equivalent simultaneous muons) and higher efficiency (below 2\,W for the full allowed supply voltage range) compared to the MPPMT front-end design tested in a prototyping phase. This was achieved by designing a dedicated power system and using high performance power supplies. Low-consumption and high performance ICs were used in all the design. The correct operation of these ICs is monitored by a slow control system based on a low sampling frequency ADC. Special care was taken in the design of the front-end including the mixed-signal PCB board, the power system and the connections with the SiPM and Back-end boards (special high-frequency connectors) to avoid digital interference and mismatch loses in the analog lines. The binary channels measurement system is based on~\cite{Hampel:2017}. This includes the SiPM power system (C11204-01 power supply), the SiPM ASIC (CITIROC) and the selected type of SiPM.

A new acquisition system based on charge measurement (ADC channels) was added to the design for enhanced dynamic range for particle detection. It is based on the measurement of the charge produced by the impinging particles in the detector and the implementation is based on summation of all analog SiPM channels by operational amplifiers. Two differential amplifiers chains provide two channels of high gain (HG) and low gain (LG). The analog to digital conversion of these two channels is handled by the dual channel ADC ADS4246. The characterization method for the ADC channels requires to obtain the average charge deposited by a muon. With this value, the conversion of the measured charge to an equivalent muon number is performed. This characterization can be applied for measurements in the laboratory and for the installed electronics at the Pierre Auger Observatory site.

Laboratory measurements and on-site data analysis show highly consistent results. The dynamic range of the detector was extended according to the analysis of the laboratory data (up to $\sim$362 equivalent simultaneous muons). The ratio of number of muons measured between identical groups of detectors is close to one and this confirms the new front-end design. All results demonstrate an adequate performance of the charge measurement system.

% created on 2020-10-08

\section*{Acknowledgments}

\begin{sloppypar}
The successful installation, commissioning, and operation of the Pierre
Auger Observatory would not have been possible without the strong
commitment and effort from the technical and administrative staff in
Malarg\"ue. We are very grateful to the following agencies and
organizations for financial support:
\end{sloppypar}

\begin{sloppypar}
Argentina -- Comisi\'on Nacional de Energ\'\i{}a At\'omica; Agencia Nacional de
Promoci\'on Cient\'\i{}fica y Tecnol\'ogica (ANPCyT); Consejo Nacional de
Investigaciones Cient\'\i{}ficas y T\'ecnicas (CONICET); Gobierno de la
Provincia de Mendoza; Municipalidad de Malarg\"ue; NDM Holdings and Valle
Las Le\~nas; in gratitude for their continuing cooperation over land
access; Australia -- the Australian Research Council; Brazil -- Conselho
Nacional de Desenvolvimento Cient\'\i{}fico e Tecnol\'ogico (CNPq);
Financiadora de Estudos e Projetos (FINEP); Funda\c{c}\~ao de Amparo \`a
Pesquisa do Estado de Rio de Janeiro (FAPERJ); S\~ao Paulo Research
Foundation (FAPESP) Grants No.~2019/10151-2, No.~2010/07359-6 and
No.~1999/05404-3; Minist\'erio da Ci\^encia, Tecnologia, Inova\c{c}\~oes e
Comunica\c{c}\~oes (MCTIC); Czech Republic -- Grant No.~MSMT CR LTT18004,
LM2015038, LM2018102, CZ.02.1.01/0.0/0.0/16{\textunderscore}013/0001402,
CZ.02.1.01/0.0/0.0/18{\textunderscore}046/0016010 and
CZ.02.1.01/0.0/0.0/17{\textunderscore}049/0008422; France -- Centre de Calcul
IN2P3/CNRS; Centre National de la Recherche Scientifique (CNRS); Conseil
R\'egional Ile-de-France; D\'epartement Physique Nucl\'eaire et Corpusculaire
(PNC-IN2P3/CNRS); D\'epartement Sciences de l'Univers (SDU-INSU/CNRS);
Institut Lagrange de Paris (ILP) Grant No.~LABEX ANR-10-LABX-63 within
the Investissements d'Avenir Programme Grant No.~ANR-11-IDEX-0004-02;
Germany -- Bundesministerium f\"ur Bildung und Forschung (BMBF); Deutsche
Forschungsgemeinschaft (DFG); Finanzministerium Baden-W\"urttemberg;
Helmholtz Alliance for Astroparticle Physics (HAP);
Helmholtz-Gemeinschaft Deutscher Forschungszentren (HGF); Ministerium
f\"ur Innovation, Wissenschaft und Forschung des Landes
Nordrhein-Westfalen; Ministerium f\"ur Wissenschaft, Forschung und Kunst
des Landes Baden-W\"urttemberg; Italy -- Istituto Nazionale di Fisica
Nucleare (INFN); Istituto Nazionale di Astrofisica (INAF); Ministero
dell'Istruzione, dell'Universit\'a e della Ricerca (MIUR); CETEMPS Center
of Excellence; Ministero degli Affari Esteri (MAE); M\'exico -- Consejo
Nacional de Ciencia y Tecnolog\'\i{}a (CONACYT) No.~167733; Universidad
Nacional Aut\'onoma de M\'exico (UNAM); PAPIIT DGAPA-UNAM; The Netherlands
-- Ministry of Education, Culture and Science; Netherlands Organisation
for Scientific Research (NWO); Dutch national e-infrastructure with the
support of SURF Cooperative; Poland -Ministry of Science and Higher
Education, grant No.~DIR/WK/2018/11; National Science Centre, Grants
No.~2013/08/M/ST9/00322, No.~2016/23/B/ST9/01635 and No.~HARMONIA
5--2013/10/M/ST9/00062, UMO-2016/22/M/ST9/00198; Portugal -- Portuguese
national funds and FEDER funds within Programa Operacional Factores de
Competitividade through Funda\c{c}\~ao para a Ci\^encia e a Tecnologia
(COMPETE); Romania -- Romanian Ministry of Education and Research, the
Program Nucleu within MCI (PN19150201/16N/2019 and PN19060102) and
project PN-III-P1-1.2-PCCDI-2017-0839/19PCCDI/2018 within PNCDI III;
Slovenia -- Slovenian Research Agency, grants P1-0031, P1-0385, I0-0033,
N1-0111; Spain -- Ministerio de Econom\'\i{}a, Industria y Competitividad
(FPA2017-85114-P and FPA2017-85197-P), Xunta de Galicia (ED431C
2017/07), Junta de Andaluc\'\i{}a (SOMM17/6104/UGR), Feder Funds, RENATA Red
Nacional Tem\'atica de Astropart\'\i{}culas (FPA2015-68783-REDT) and Mar\'\i{}a de
Maeztu Unit of Excellence (MDM-2016-0692); USA -- Department of Energy,
Contracts No.~DE-AC02-07CH11359, No.~DE-FR02-04ER41300,
No.~DE-FG02-99ER41107 and No.~DE-SC0011689; National Science Foundation,
Grant No.~0450696; The Grainger Foundation; Marie Curie-IRSES/EPLANET;
European Particle Physics Latin American Network; and UNESCO.
\end{sloppypar}

\newpage
\appendix

\section{Schematics}
\label{appendix_schematics}

\subsection{CITIROC ASIC}
\begin{figure}[H]
	\centering
	\includegraphics[height=\textwidth,width=0.83\textheight,angle=90]{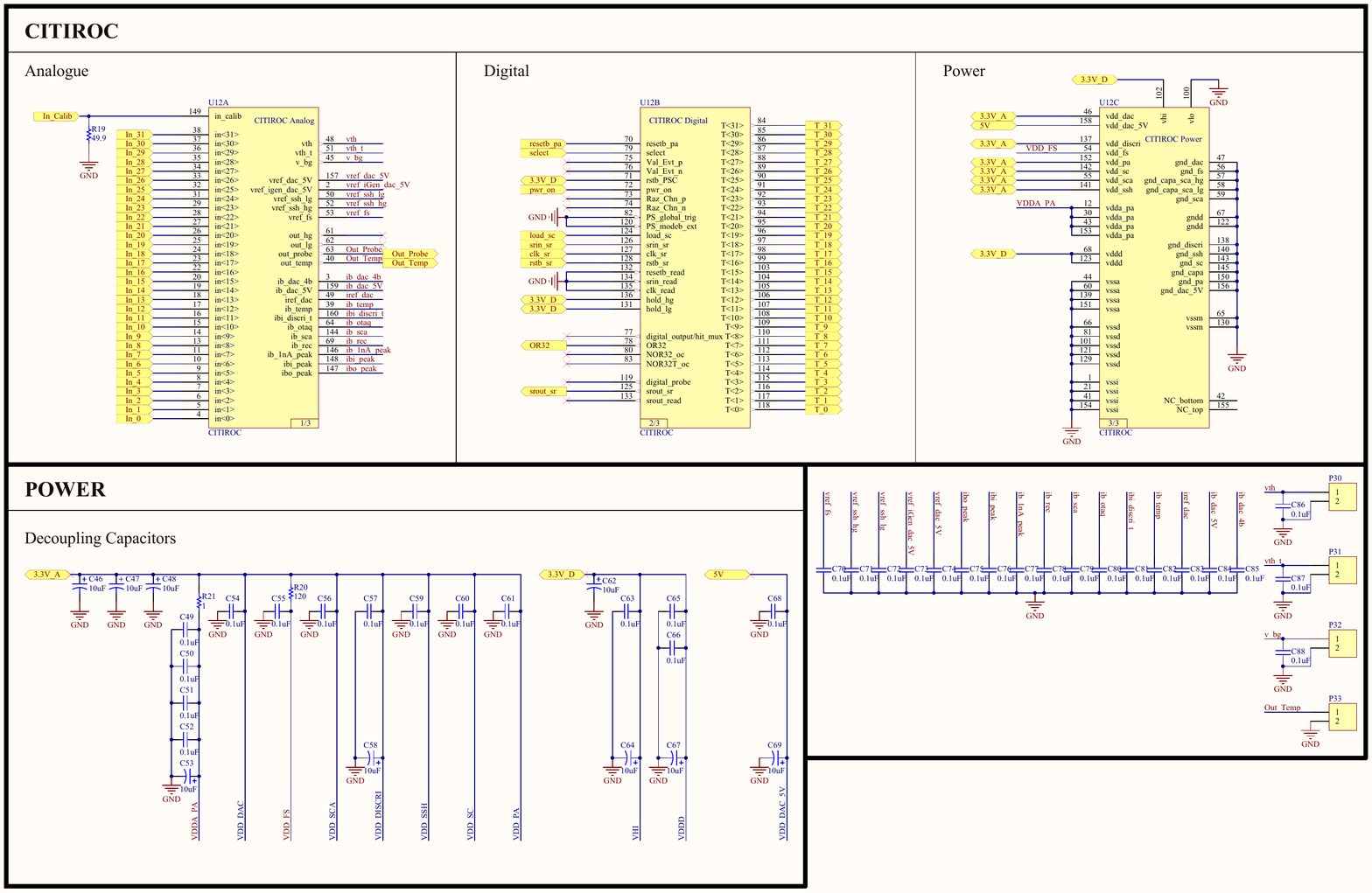}	
	\caption{Interfacing the CITIROC ASIC. The schematics is divided into 3 different parts: the analog part (all SiPM anodes), the digital part (digital outputs to back-end electronics) and the power supply part (including the power supply capacitors).}
	\label{fig:citiroc_sche}
\end{figure}

\subsection{Non-isolated 5\,V power supply}

\begin{figure}[H]
	\centering
	\includegraphics[width=.8\textwidth]{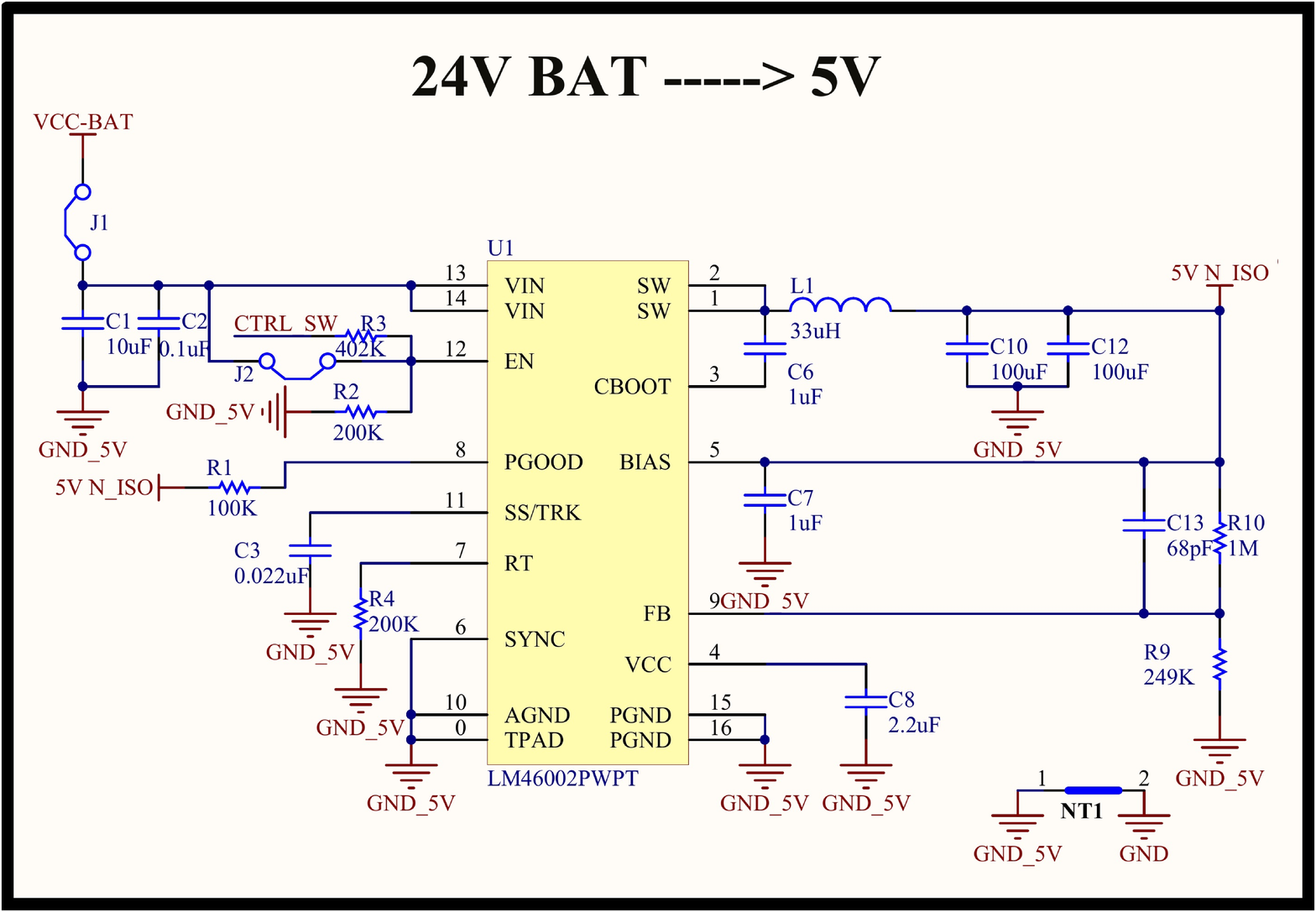}
	\caption{Schematics for the non-isolated 5\,V power supply. The LM46002 regulator is a synchronous step-down DC/DC converter capable of driving a load currents of up to 2\,A from an input voltage ranging from 3.5\,V to 60\,V. This option is appropriate in a photovoltaic system where the input voltage from battery is limited to a maximum of 32\,V at full solar illumination. The CTRL\_SW signal controls the power on/off of the converter.}
	\label{fig:non-isolated_supply}
\end{figure}

\subsection{Isolated 5\,V power supply}

\begin{figure}[H]
	\centering
	\includegraphics[width=.7\textwidth]{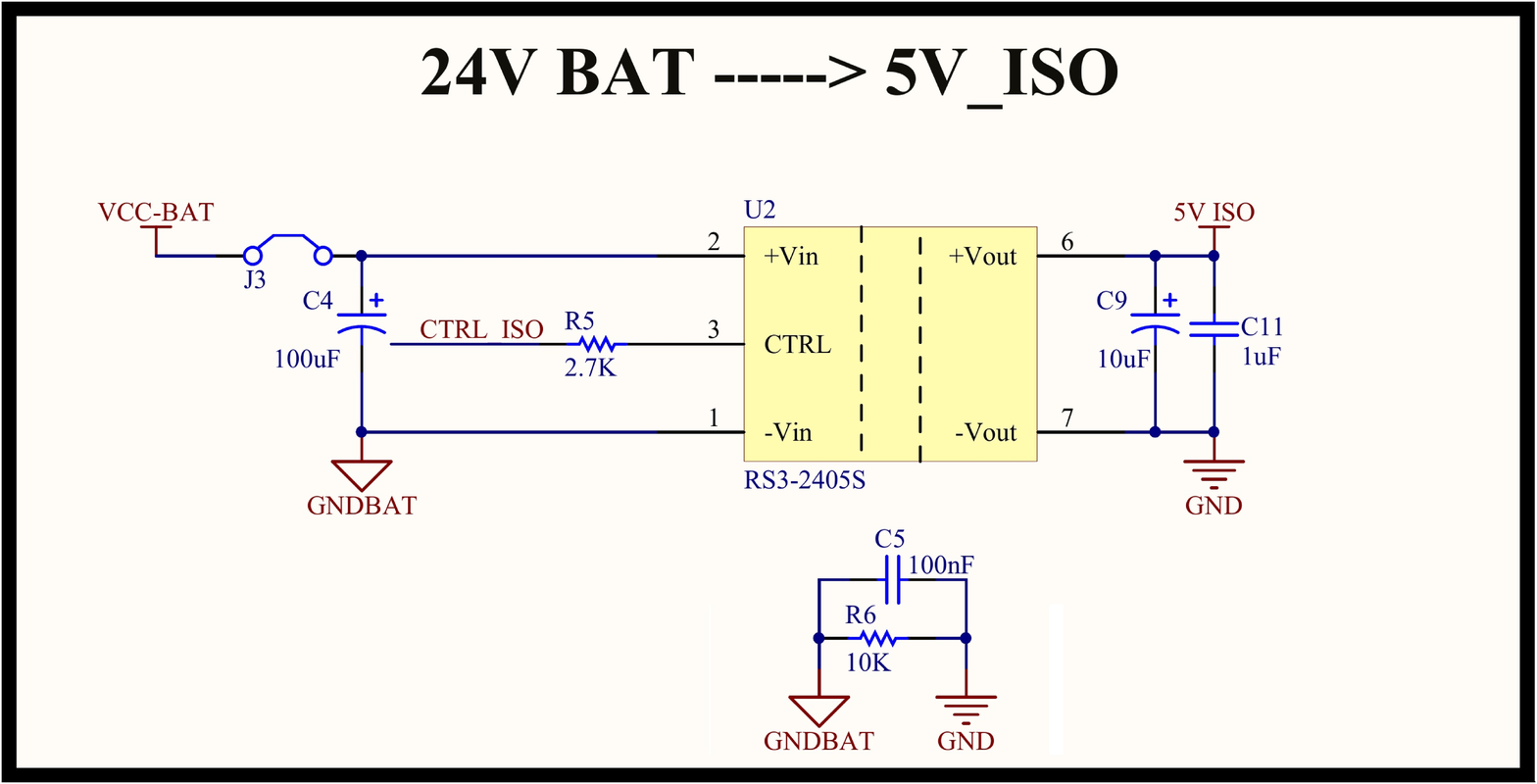}
	\caption{Schematics of the isolated 5\,V power supply. The isolated DC/DC converter RS3-2405S is capable of driving up to 600\,mA (3\,W maximum output power) from an input voltage ranging from 18\,V to 36\,V. The galvanic isolation avoids ground loops. Capacitor C5 is needed to reduce the output voltage ripple and resistor R6 avoids problems with electrostatic discharge. The CTRL\_ISO signal controls the power on/off of the converter.}
	\label{fig:isolated_supply}
\end{figure}

\subsection{SiPM power supply}
\begin{figure}[H]
	\centering
	\includegraphics[width=.75\textwidth]{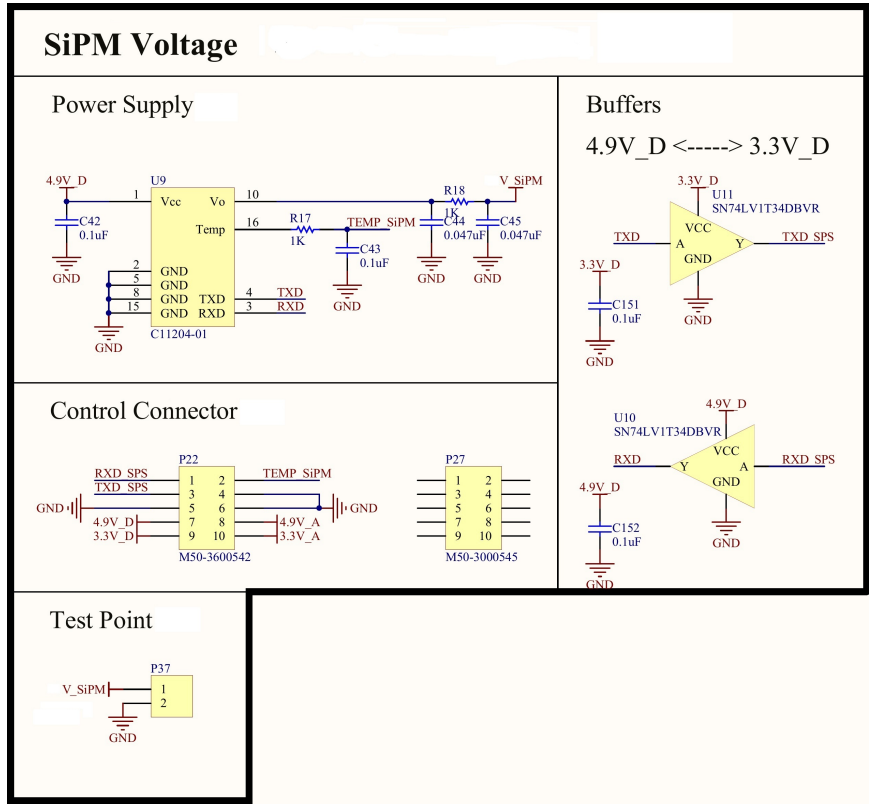}
	\caption{SiPM power supply schematics. The SN74LV1T34~\cite{DS-SN74LV1T34} is a single supply buffer logic level shifter used to adapt the signal voltage levels of the 5\,V SiPM power supply and the voltage levels of the back-end of 3.3\,V.}
	\label{fig:hv_power_supply}
\end{figure}

\subsection{Back-end power connector}
\begin{figure}[H]
	\centering
	\includegraphics[width=.7\textwidth]{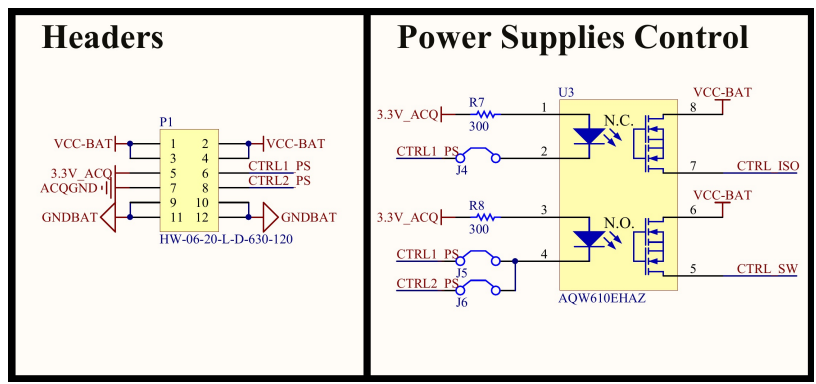}
	\caption{Back-end power connector schematics. The connector P1 is a high-temp Flexible board stacker to provide the interface between the front-end and back-end boards. The back-end electronics controls the power on/off of the two converters explained in section~\ref{subsec:Power system} through optocouplers (right side) to avoid possible ground loops (galvanic isolation).}
	\label{fig:power_connector}
\end{figure}

\subsection{Slow control monitor system}
\begin{figure}[H]
	\centering
	\includegraphics[width=.7\textwidth]{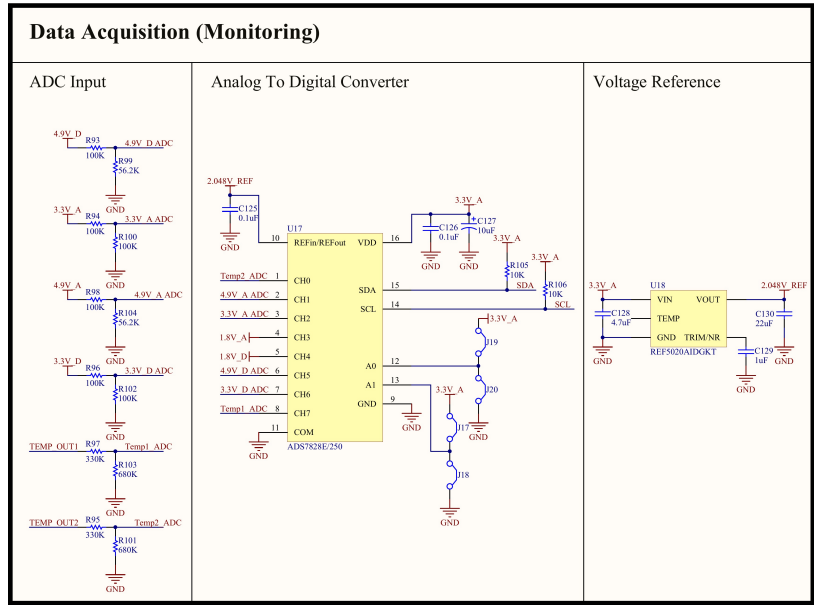}
	\caption{Slow control monitor system. The low power ADC measures all supply voltages of the board and the two temperature voltage outputs of both CITIROCs. A resistor divider at the ADC input adapts the range of the monitored voltages to the ADC input range given by an external 2.048\,V voltage reference.}
	\label{fig:slow_monitor}
\end{figure}

\subsection{Adder amplifiers}
\begin{figure}[H]
	\centering
	\includegraphics[width=.7\textwidth]{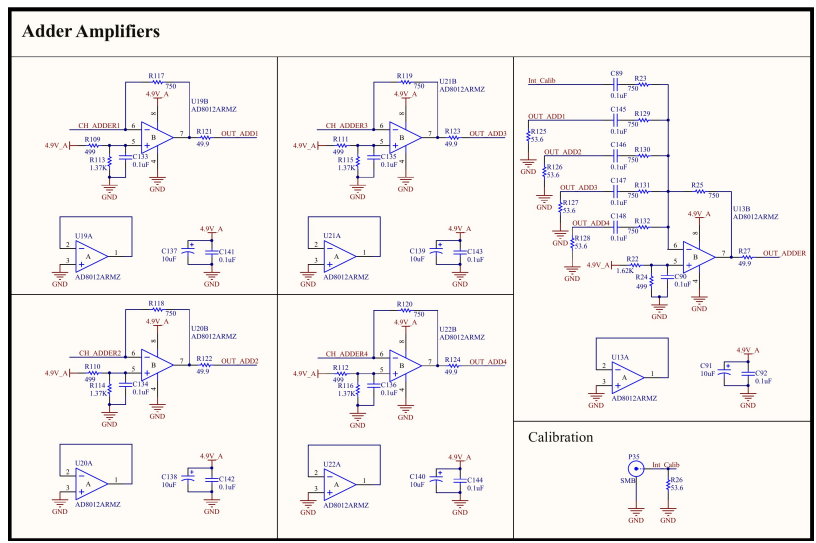}
	\caption{Adder amplifiers schematics. The four amplifiers on the left side (U19, U20, U21 and U22) add up the 64 SiPM anode signals in groups of 16 channels. The amplifier on the right side (U13) is the sum of the outputs of the four previous amplifiers. This amplifier provides a calibration input.}
	\label{fig:adder_amplifiers}
\end{figure}

\subsection{Differential output amplifiers}
\begin{figure}[H]
	\centering
	\includegraphics[width=.9\textwidth]{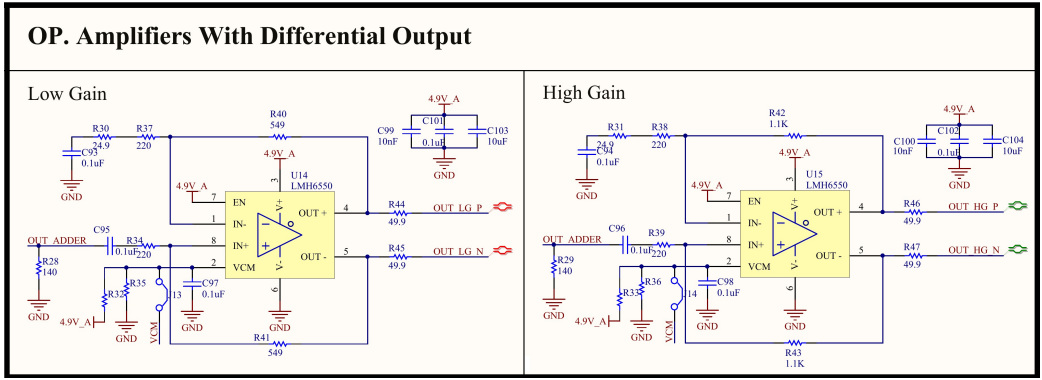}
	\caption{Differential output amplifiers schematics. At this stage, the ADC channels is divided into a high gain channel (right side amplifier) and a low gain channel (left side amplifier) branch.}
	\label{fig:opamp_diff_output}
\end{figure}

\subsection{Analog to digital converter}
\begin{figure}[H]
	\centering
	\includegraphics[width=.9\textwidth]{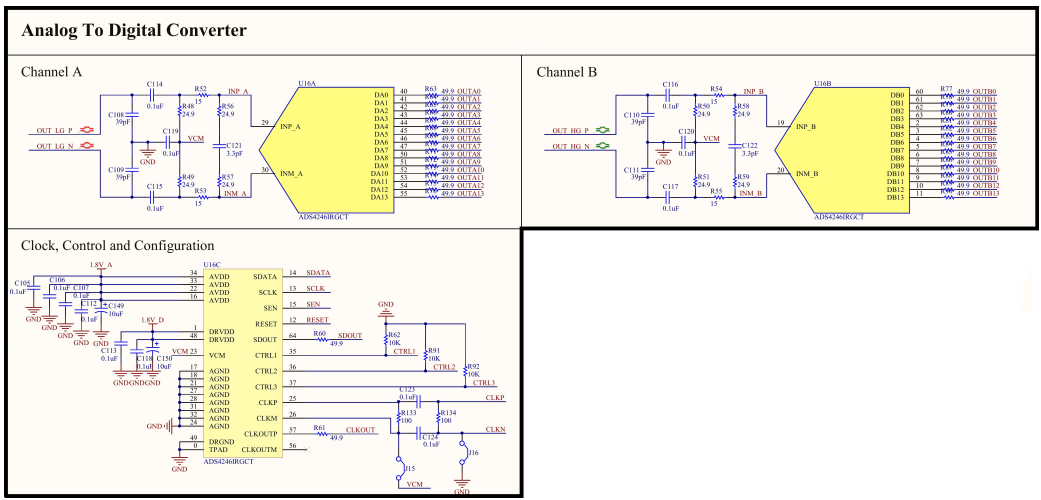}
	\caption{Analog to digital converter schematics. Each channels (high and low gain) is connected to one input of the ADC after passing through an anti-aliasing filter.	The serial 49.9\,$\Omega$ resistors on the channel outputs limit the output current at high capacitive loads.}
	\label{fig:adc_sche}
\end{figure}

\bibliography{Datasheets.bib,Miscelaneas.bib,Papers.bib,Proceeding.bib,Reportes.bib,Tesis.bib}

\begin{center}
\rule{0.1\columnwidth}{0.5pt}\,\raisebox{-0.5pt}{\rule{0.05\columnwidth}{1.5pt}}~\raisebox{-0.375ex}{\scriptsize$\bullet$}~\raisebox{-0.5pt}{\rule{0.05\columnwidth}{1.5pt}}\,\rule{0.1\columnwidth}{0.5pt}
\end{center}

\newpage
\section*{The Pierre Auger Collaboration}
\addcontentsline{toc}{section}{The Pierre Auger Collaboration}
{\small
% created on 2020-10-08

A.~Aab$^{76}$,
P.~Abreu$^{68}$,
M.~Aglietta$^{51,50}$,
J.M.~Albury$^{12}$,
I.~Allekotte$^{1}$,
A.~Almela$^{8,11}$,
J.~Alvarez-Mu\~niz$^{75}$,
R.~Alves Batista$^{76}$,
G.A.~Anastasi$^{59,50}$,
L.~Anchordoqui$^{83}$
% ...

% created on 2020-10-08

% needs \usepackage{enumitem}
\begin{description}[labelsep=0.2em,align=right,labelwidth=0.7em,labelindent=0em,leftmargin=2em,noitemsep]
\item[$^{1}$] Centro At\'omico Bariloche and Instituto Balseiro (CNEA-UNCuyo-CONICET), San Carlos de Bariloche, Argentina
\item[$^{2}$] Centro de Investigaciones en L\'aseres y Aplicaciones, CITEDEF and CONICET, Villa Martelli, Argentina
\item[$^{3}$] Departamento de F\'\i{}sica and Departamento de Ciencias de la Atm\'osfera y los Oc\'eanos, FCEyN, Universidad de Buenos Aires and CONICET, Buenos Aires, Argentina
\item[$^{4}$] IFLP, Universidad Nacional de La Plata and CONICET, La Plata, Argentina
\item[$^{5}$] Instituto de Astronom\'\i{}a y F\'\i{}sica del Espacio (IAFE, CONICET-UBA), Buenos Aires, Argentina
\item[$^{6}$] Instituto de F\'\i{}sica de Rosario (IFIR) -- CONICET/U.N.R.\ and Facultad de Ciencias Bioqu\'\i{}micas y Farmac\'euticas U.N.R., Rosario, Argentina
\item[$^{7}$] Instituto de Tecnolog\'\i{}as en Detecci\'on y Astropart\'\i{}culas (CNEA, CONICET, UNSAM), and Universidad Tecnol\'ogica Nacional -- Facultad Regional Mendoza (CONICET/CNEA), Mendoza, Argentina
\item[$^{8}$] Instituto de Tecnolog\'\i{}as en Detecci\'on y Astropart\'\i{}culas (CNEA, CONICET, UNSAM), Buenos Aires, Argentina
\item[$^{9}$] Observatorio Pierre Auger, Malarg\"ue, Argentina
\item[$^{10}$] Observatorio Pierre Auger and Comisi\'on Nacional de Energ\'\i{}a At\'omica, Malarg\"ue, Argentina
\item[$^{11}$] Universidad Tecnol\'ogica Nacional -- Facultad Regional Buenos Aires, Buenos Aires, Argentina
\item[$^{12}$] University of Adelaide, Adelaide, S.A., Australia
\item[$^{13}$] Universit\'e Libre de Bruxelles (ULB), Brussels, Belgium
\item[$^{14}$] Vrije Universiteit Brussels, Brussels, Belgium
\item[$^{15}$] Centro Brasileiro de Pesquisas Fisicas, Rio de Janeiro, RJ, Brazil
\item[$^{16}$] Centro Federal de Educa\c{c}\~ao Tecnol\'ogica Celso Suckow da Fonseca, Nova Friburgo, Brazil
\item[$^{17}$] Universidade de S\~ao Paulo, Escola de Engenharia de Lorena, Lorena, SP, Brazil
\item[$^{18}$] Universidade de S\~ao Paulo, Instituto de F\'\i{}sica de S\~ao Carlos, S\~ao Carlos, SP, Brazil
\item[$^{19}$] Universidade de S\~ao Paulo, Instituto de F\'\i{}sica, S\~ao Paulo, SP, Brazil
\item[$^{20}$] Universidade Estadual de Campinas, IFGW, Campinas, SP, Brazil
\item[$^{21}$] Universidade Estadual de Feira de Santana, Feira de Santana, Brazil
\item[$^{22}$] Universidade Federal do ABC, Santo Andr\'e, SP, Brazil
\item[$^{23}$] Universidade Federal do Paran\'a, Setor Palotina, Palotina, Brazil
\item[$^{24}$] Universidade Federal do Rio de Janeiro, Instituto de F\'\i{}sica, Rio de Janeiro, RJ, Brazil
\item[$^{25}$] Universidade Federal do Rio de Janeiro (UFRJ), Observat\'orio do Valongo, Rio de Janeiro, RJ, Brazil
\item[$^{26}$] Universidade Federal Fluminense, EEIMVR, Volta Redonda, RJ, Brazil
\item[$^{27}$] Universidad de Medell\'\i{}n, Medell\'\i{}n, Colombia
\item[$^{28}$] Universidad Industrial de Santander, Bucaramanga, Colombia
\item[$^{29}$] Charles University, Faculty of Mathematics and Physics, Institute of Particle and Nuclear Physics, Prague, Czech Republic
\item[$^{30}$] Institute of Physics of the Czech Academy of Sciences, Prague, Czech Republic
\item[$^{31}$] Palacky University, RCPTM, Olomouc, Czech Republic
\item[$^{32}$] CNRS/IN2P3, IJCLab, Universit\'e Paris-Saclay, Orsay, France
\item[$^{33}$] Laboratoire de Physique Nucl\'eaire et de Hautes Energies (LPNHE), Sorbonne Universit\'e, Universit\'e de Paris, CNRS-IN2P3, Paris, France
\item[$^{34}$] Univ.\ Grenoble Alpes, CNRS, Grenoble Institute of Engineering Univ.\ Grenoble Alpes, LPSC-IN2P3, 38000 Grenoble, France
\item[$^{35}$] Universit\'e Paris-Saclay, CNRS/IN2P3, IJCLab, Orsay, France
\item[$^{36}$] Bergische Universit\"at Wuppertal, Department of Physics, Wuppertal, Germany
\item[$^{37}$] Karlsruhe Institute of Technology, Institute for Experimental Particle Physics (ETP), Karlsruhe, Germany
\item[$^{38}$] Karlsruhe Institute of Technology, Institute for Astroparticle Physics, Karlsruhe, Germany
\item[$^{39}$] Karlsruhe Institute of Technology, Institut f\"ur Prozessdatenverarbeitung und Elektronik, Karlsruhe, Germany
\item[$^{40}$] RWTH Aachen University, III.\ Physikalisches Institut A, Aachen, Germany
\item[$^{41}$] Universit\"at Hamburg, II.\ Institut f\"ur Theoretische Physik, Hamburg, Germany
\item[$^{42}$] Universit\"at Siegen, Department Physik -- Experimentelle Teilchenphysik, Siegen, Germany
\item[$^{43}$] Gran Sasso Science Institute, L'Aquila, Italy
\item[$^{44}$] INFN Laboratori Nazionali del Gran Sasso, Assergi (L'Aquila), Italy
\item[$^{45}$] INFN, Sezione di Catania, Catania, Italy
\item[$^{46}$] INFN, Sezione di Lecce, Lecce, Italy
\item[$^{47}$] INFN, Sezione di Milano, Milano, Italy
\item[$^{48}$] INFN, Sezione di Napoli, Napoli, Italy
\item[$^{49}$] INFN, Sezione di Roma ``Tor Vergata'', Roma, Italy
\item[$^{50}$] INFN, Sezione di Torino, Torino, Italy
\item[$^{51}$] Osservatorio Astrofisico di Torino (INAF), Torino, Italy
\item[$^{52}$] Politecnico di Milano, Dipartimento di Scienze e Tecnologie Aerospaziali , Milano, Italy
\item[$^{53}$] Universit\`a del Salento, Dipartimento di Matematica e Fisica ``E.\ De Giorgi'', Lecce, Italy
\item[$^{54}$] Universit\`a dell'Aquila, Dipartimento di Scienze Fisiche e Chimiche, L'Aquila, Italy
\item[$^{55}$] Universit\`a di Catania, Dipartimento di Fisica e Astronomia, Catania, Italy
\item[$^{56}$] Universit\`a di Milano, Dipartimento di Fisica, Milano, Italy
\item[$^{57}$] Universit\`a di Napoli ``Federico II'', Dipartimento di Fisica ``Ettore Pancini'', Napoli, Italy
\item[$^{58}$] Universit\`a di Roma ``Tor Vergata'', Dipartimento di Fisica, Roma, Italy
\item[$^{59}$] Universit\`a Torino, Dipartimento di Fisica, Torino, Italy
\item[$^{60}$] Benem\'erita Universidad Aut\'onoma de Puebla, Puebla, M\'exico
\item[$^{61}$] Centro de Investigaci\'on y de Estudios Avanzados del IPN (CINVESTAV), M\'exico, D.F., M\'exico
\item[$^{62}$] Unidad Profesional Interdisciplinaria en Ingenier\'\i{}a y Tecnolog\'\i{}as Avanzadas del Instituto Polit\'ecnico Nacional (UPIITA-IPN), M\'exico, D.F., M\'exico
\item[$^{63}$] Universidad Aut\'onoma de Chiapas, Tuxtla Guti\'errez, Chiapas, M\'exico
\item[$^{64}$] Universidad Nacional Aut\'onoma de M\'exico, M\'exico, D.F., M\'exico
\item[$^{65}$] Institute of Nuclear Physics PAN, Krakow, Poland
\item[$^{66}$] University of \L{}\'od\'z, Faculty of Astrophysics, \L{}\'od\'z, Poland
\item[$^{67}$] University of \L{}\'od\'z, Faculty of High-Energy Astrophysics,\L{}\'od\'z, Poland
\item[$^{68}$] Laborat\'orio de Instrumenta\c{c}\~ao e F\'\i{}sica Experimental de Part\'\i{}culas -- LIP and Instituto Superior T\'ecnico -- IST, Universidade de Lisboa -- UL, Lisboa, Portugal
\item[$^{69}$] ``Horia Hulubei'' National Institute for Physics and Nuclear Engineering, Bucharest-Magurele, Romania
\item[$^{70}$] Institute of Space Science, Bucharest-Magurele, Romania
\item[$^{71}$] University Politehnica of Bucharest, Bucharest, Romania
\item[$^{72}$] Center for Astrophysics and Cosmology (CAC), University of Nova Gorica, Nova Gorica, Slovenia
\item[$^{73}$] Experimental Particle Physics Department, J.\ Stefan Institute, Ljubljana, Slovenia
\item[$^{74}$] Universidad de Granada and C.A.F.P.E., Granada, Spain
\item[$^{75}$] Instituto Galego de F\'\i{}sica de Altas Enerx\'\i{}as (IGFAE), Universidade de Santiago de Compostela, Santiago de Compostela, Spain
\item[$^{76}$] IMAPP, Radboud University Nijmegen, Nijmegen, The Netherlands
\item[$^{77}$] KVI -- Center for Advanced Radiation Technology, University of Groningen, Groningen, The Netherlands
\item[$^{78}$] Nationaal Instituut voor Kernfysica en Hoge Energie Fysica (NIKHEF), Science Park, Amsterdam, The Netherlands
\item[$^{79}$] Stichting Astronomisch Onderzoek in Nederland (ASTRON), Dwingeloo, The Netherlands
\item[$^{80}$] Universiteit van Amsterdam, Faculty of Science, Amsterdam, The Netherlands
\item[$^{81}$] Case Western Reserve University, Cleveland, OH, USA
\item[$^{82}$] Colorado School of Mines, Golden, CO, USA
\item[$^{83}$] Department of Physics and Astronomy, Lehman College, City University of New York, Bronx, NY, USA
\item[$^{84}$] Louisiana State University, Baton Rouge, LA, USA
\item[$^{85}$] Michigan Technological University, Houghton, MI, USA
\item[$^{86}$] New York University, New York, NY, USA
\item[$^{87}$] Pennsylvania State University, University Park, PA, USA
\item[$^{88}$] University of Chicago, Enrico Fermi Institute, Chicago, IL, USA
\item[$^{89}$] University of Delaware, Department of Physics and Astronomy, Bartol Research Institute, Newark, DE, USA
\item[] -----
\item[$^{a}$] Fermi National Accelerator Laboratory, USA
\item[$^{b}$] also at Universidade Federal de Alfenas, Po\c{c}os de Caldas, Brazil
\item[$^{c}$] Max-Planck-Institut f\"ur Radioastronomie, Bonn, Germany
\item[$^{d}$] School of Physics and Astronomy, University of Leeds, Leeds, United Kingdom
\item[$^{e}$] Fermi National Accelerator Laboratory, Fermilab, Batavia, IL, USA
\item[$^{f}$] also at Karlsruhe Institute of Technology, Karlsruhe, Germany
\item[$^{g}$] Colorado State University, Fort Collins, CO, USA
\item[$^{h}$] now at Hakubi Center for Advanced Research and Graduate School of Science, Kyoto University, Kyoto, Japan
\item[$^{i}$] also at University of Bucharest, Physics Department, Bucharest, Romania
\end{description}

}
\end{document}